\documentclass[sigconf]{acmart}

\AtBeginDocument{%
  }

\copyrightyear{2025}
\acmYear{2025}
\setcopyright{cc}
\setcctype{by}
\acmConference[IMC '25]{Proceedings of the 2025 ACM Internet Measurement
Conference}{October 28--31, 2025}{Madison, WI, USA}
\acmBooktitle{Proceedings of the 2025 ACM Internet Measurement Conference (IMC '25), October 28--31, 2025, Madison, WI, USA}
\acmDOI{10.1145/3730567.3732917}
\acmISBN{979-8-4007-1860-1/2025/10}

\newcommand{\ie}{i.e., \@}
\newcommand{\eg}{e.g., \@}

\newcommand{\etal}{et al.\xspace}

\newcommand{\parx}[1]{\noindent\textbf{#1}\xspace}

\usepackage{xspace}
\usepackage{cleveref}
\usepackage{graphicx}
\usepackage{algorithm}
\usepackage[noend]{algpseudocode}
\usepackage{balance}

\usepackage{enumitem}
\usepackage{mdwlist}
\setlist{nolistsep}
\setlist[description]{noitemsep,topsep=0pt,parsep=0pt,partopsep=0pt,leftmargin=0pt}

\usepackage[absolute,showboxes]{textpos}

\begin{document}

\settopmatter{printfolios=true}
\setlength{\TPHorizModule}{\paperwidth}
\setlength{\TPVertModule}{\paperheight}
\TPMargin{5pt}
\begin{textblock}{0.8}(0.1,0.02)
     \noindent
     \footnotesize
     If you cite this paper, please use the IMC 2025 reference:
     Fariba Osali, Khwaja Zubair Sediqi, and Oliver Gasser. 2025. Sibling Prefixes:
Identifying Similarities in IPv4 and IPv6 Prefixes. In \textit{Proceedings of the
2025 ACM Internet Measurement Conference (IMC ’25), October 28–31, 2025,
Madison, WI, USA.} ACM, New York, NY, USA, 20 pages. https://doi.org/10.
1145/3730567.3732917
\end{textblock}

\title[Sibling Prefixes: Identifying Similarities in IPv4 and IPv6 Prefixes]{Sibling Prefixes: \\Identifying Similarities in IPv4 and IPv6 Prefixes}

\author{Fariba Osali}
\affiliation{%
  \institution{Max Planck Institute for Informatics}
  \city{Saarbrücken}
  \country{Germany}
  }
\email{fosali@mpi-inf.mpg.de}

\author{Khwaja Zubair Sediqi}
\affiliation{%
  \institution{Max Planck Institute for Informatics, Saarland University}
  \city{Saarbrücken}
  \country{Germany}
  }
\email{zsediqi@mpi-inf.mpg.de}

\author{Oliver Gasser}
\affiliation{%
  \institution{IPinfo}
  \city{Seattle}
  \country{USA}
  }
\email{oliver@ipinfo.io}

\renewcommand{\shortauthors}{Fariba Osali, Khwaja Zubair Sediqi, Oliver Gasser}

\begin{abstract}
Since the standardization of IPv6 in 1998, both versions of the Internet Protocol have coexisted in the Internet.
Clients usually run algorithms such as Happy Eyeballs, to decide whether to connect to an IPv4 or IPv6 endpoint for dual-stack domains.
To identify whether two addresses belong to the same device or service, researchers have proposed different forms of alias resolution techniques.
Similarly, one can also form siblings of IPv4 and IPv6 addresses belonging to the same device.
Traditionally, all of these approaches have focused on individual IP addresses.

In this work, we propose the concept of ``sibling prefixes'', where we extend the definition of an IPv4-IPv6 sibling to two IP prefixes---one IPv4 prefix and its sibling IPv6 prefix.
We present a technique based on large-scale DNS resolution data to identify 76k IPv4-IPv6 sibling prefixes. We find sibling prefixes to be relatively stable over time.
We present SP-Tuner algorithm to tune the CIDR size of sibling prefixes and improve the perfect match siblings from 52\% to 82\%.
For more than half of sibling prefixes, the organization names for their IPv4 and IPv6 origin ASes are identical, and 60\% of all sibling prefixes have at least one of the prefixes with a valid ROV status in RPKI. Furthermore, we identify sibling prefixes in 24 hypergiant and CDN networks. 
Finally, we plan to regularly publish a list of sibling prefixes to be used by network operators and fellow researchers in dual-stack studies.

\end{abstract}



\maketitle

\section{Introduction}
\label{sec:introduction}

Since the beginnings of the standardization of IPv6 in 1998 \cite{rfc1883}, the traditional IPv4 Internet and the newer IPv6 Internet have existed side-by-side.
Initial deployment of IPv6-enabled services as well as clients has been slow, taking up to 2013 to reach 1\% of Internet traffic \cite{czyz2014measuring}.
With the exhaustion of the IPv4 address space \cite{richter2015primer,prehn2020wells} and the preference for IPv6 over IPv4 in connection establishments \cite{rfc6555,rfc8305}, IPv6 sees continuous growth, with more than 219,000 IPv6 prefixes in the global routing table in November 2024 \cite{potaroo_v6}, more than 75\% of top 1k popular domains being accessible over IPv6 in August 2022 \cite{streibelt2023dnsipv6}, and more than 45\% of clients accessing Google via IPv6 in November 2024 \cite{google_v6}.
This continuing growth in IPv6 requires a better understanding of the IPv4 and IPv6 interdependencies.
One such interdependency is siblings, \ie a connection between an IPv4 and IPv6 address belonging to the same host.
The identification of such connections, \ie sibling detection has been widely studied in the past \cite{berger2013internet,beverly2015server,scheitle2017large,albakour2021third,albakour2023pushing}.
To date, however, siblings have only been analyzed on an IP address level.

In this paper, we strive to push the boundaries of sibling detection and extend their reach from IP addresses to IP prefixes.
We introduce the concept of a \emph{sibling prefix} as an IPv4 and IPv6 prefix pair with similar services (\ie IPv4 and IPv6 prefixes sharing similar domain names).
We present a technique to detect these sibling prefixes by leveraging large-scale DNS resolution results.
Identifying sibling prefixes allows us to better understand interdependencies between the IPv4 and IPv6 Internet in several ways and is the first step towards a more comprehensive understanding of differences in network performance, policy, geolocation 
and security posture of services running concurrently on both IPv4 and IPv6.
Sibling prefixes can show shared networking infrastructure, highlight the presence of backup paths (or lack thereof) between IPv4 and IPv6, and shed light on the deployment and management of dual-stack services in different networks.
Moreover, identification of sibling prefixes can help network operators to make more informed and inclusive routing decision, by considering both IPv4 and IPv6 as part of sibling prefixes.
The inconsistent treatment of IPv4 and IPv6 traffic, especially in dual-stack environments, might lead to operational inefficiency or security vulnerabilities. Identifying sibling prefixes can thus help network operators apply consistent networking policies across both IP address families.
For instance, network operators might want to prioritize, filter, or block traffic/domains of IPv4 prefixes, and identified sibling prefixes allows to do this for the IPv6 counterpart as well.
Furthermore, service operators can leverage sibling prefixes to apply knowledge learned about a prefix for one IP version to the other IP version.
One example are geolocation database providers using sibling prefixes to transfer geolocation information from IPv4 to IPv6 using sibling prefixes, thus improving geolocation across IP version boundaries.
Our main contributions in this paper are as follows:

\begin{figure*}[t] %
    \begin{minipage}[t]{0.66\textwidth}
      \centering
      \includegraphics[width=\linewidth]{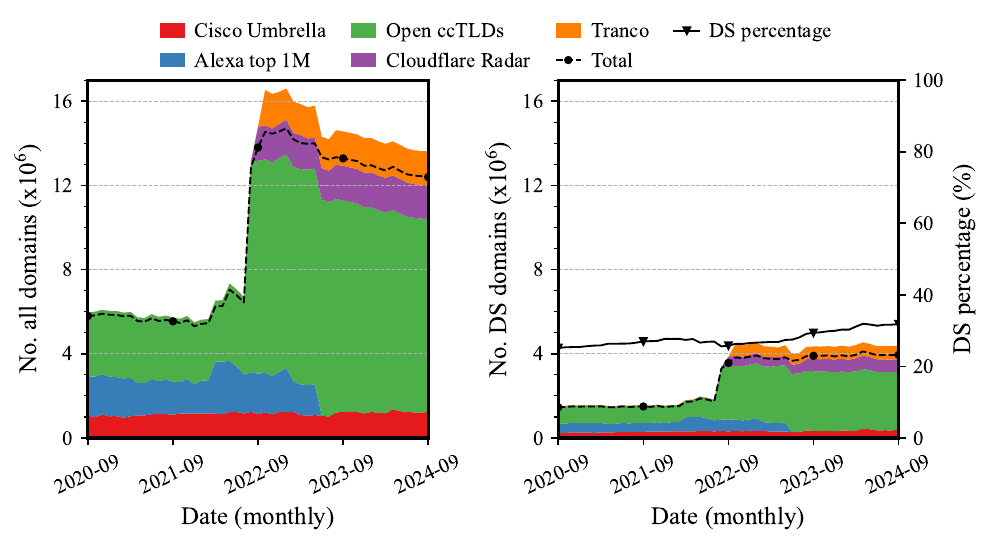}
      \caption{Number of all domains (left) and dual-stack (DS) domains (right) over time in the OpenINTEL dataset.}
  
      \label{fig:ds_overtime}
    \end{minipage}
    \hfill 
    \begin{minipage}[t]{0.32\textwidth}
      \centering
      \includegraphics[width=\linewidth]{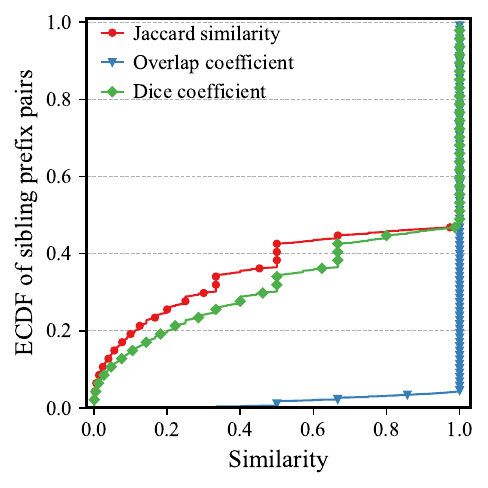}
      \caption{Comparison of Jaccard, Dice, and overlap coefficient similarity metrics.}
      \label{fig:Jaccard_vs_Dice_vs_overlap_0day}
    \end{minipage}
  \end{figure*}

\begin{itemize}[leftmargin=*]
    \item \textbf{Sibling prefix detection methodology:}
    We present a novel technique to detect sibling prefixes and thoroughly evaluate their suitability.
    We validate the sibling prefix relationship using the 2200 RIPE Atlas probes, 260 globally distributed dual-stack virtual private servers, and data from port scan results, see \Cref{sec:method}.

    \item \textbf{Tuning sibling prefix sizes:}
        We develop the sibling prefix tuner (SP-Tuner) algorithm that searches and derives new pairs of sibling prefixes from existing ones with a higher Jaccard similarity, by looking for more suitable CIDR sizes than the ones observed in Internet routing.
        The SP-Tuner algorithm increases the percentage of sibling prefixes with Jaccard similarity value of 1 (\ie perfect match siblings) from 52\% to 82\%, see \Cref{sec:sp_tuner,subsec:sp-tuner_impact}.
        We publish data and code of our detection methodology at \href{https://sibling-prefixes.github.io}{sibling-prefixes.github.io}.

      \item \textbf{Large-scale sibling prefix analysis:}
      We perform large-scale sibling prefix measurements using data from IPv4 and IPv6 DNS resolutions, finding around 46k unique IPv4 and almost 39k unique IPv6 prefixes, resulting in more than 76k sibling prefix pairs, of which more than half share the same organization names for their IPv4 and IPv6 origin ASes.
      We also conduct longitudinal measurements and find that sibling prefixes are relatively stable over time.
      While IT organizations have the highest number of sibling prefixes, we further identify sibling prefixes in 24 hypergiants and CDNs.
      Moreover, we use RPKI data to analyze sibling prefixes' route origin validation (ROV) status, finding over 60\% of sibling prefix pairs having at least one valid origin AS, see \Cref{sec:sibling_prefixes}.

      \item \textbf{Impact of sibling prefixes:}
          We discuss the potential uses and impact of sibling prefixes, highlighting the need for tailored prefix size choices, and drawing attention to a paradigm shift from IP addresses to domains.
          Finally, we plan to regularly publish a list of sibling prefixes for use by fellow researchers and network operators, see \Cref{sec:discussion}.
    \end{itemize}

\section{Datasets}
\label{sec:dataset}

This section provides details on datasets we use for identifying and characterizing sibling prefixes.

\subsection{DNS Dataset}
\label{subsec:dns_dataset}

One of the cornerstones of our methodology to detect sibling prefixes relies on shared domain names between IPv4 and IPv6 prefixes.
We leverage domain names from the OpenINTEL DNS dataset \cite{openintel} to obtain DNS resolution results from large-scale measurements.
The OpenINTEL dataset consists of DNS resolution results of the Alexa top 1M \cite{alexa}, Cloudflare Radar \cite{radar},  Tranco \cite{tranco}, Cisco Umbrella \cite{umbrella}, and open ccTLD domains.
We collect four years of OpenINTEL data on every second Wednesday of each month from September 2020 to September 2024, resulting in 49 snapshots.

\Cref{fig:ds_overtime} shows the longitudinal evolution of the OpenINTEL dataset over time.
The left subplot shows the overall number of domains in the dataset, whereas the right subplot focuses on dual-stack (DS) domains DS domains are the relevant subset of domains for identifying sibling prefixes, as we rely on shared IPv4-IPv6 domains within sibling pairs. 
As can be seen, the overall number of domains as well as DS domains in the dataset has been growing over time. 
This growth is mainly due to additional datasets being added to OpenINTEL, such as Tranco in September 2022 and Cloudflare Radar October 2022.
The largest increase in the total number of domains occurs in August 2022, which corresponds to the inclusion of ``.fr'' TLD domains 6.35M when it was added to the set of open ccTLDs \cite{sommese2024local}.
On the contrary, the removal of the Alexa top 1M dataset in May 2023, leads to a slight decrease in the number of domains.
Furthermore, the percentage of DS domains is also slightly increasing over time from 25.2\% in September 2020 to 31.8\% in September 2024. 
This highlights the increasing deployment of dual-stack domains and, therefore, the need to identify these deployments using sibling prefixes.
In the remainder of the paper, we leverage 3.95 million unique DS domains to identify sibling prefixes.
\subsection{IP to Prefix and AS Dataset}
\label{subsec:ip-and-as-number-dataset}
The OpenINTEL dataset provides prefix and AS information for A and AAAA DNS records, but around 1\% of records lack prefix or AS information. To address this, we use data from the Routeviews project \cite{routeviews} to identify the missing prefix and AS information.
For less than 0.01\% of DS domains, we observe private, invalid, or reserved IP addresses, which we discard and do not consider for further analysis. 
The latest snapshot of the OpenINTEL dataset (September 11, 2024) contains 271.5k IPv4 addresses and 978.4k IPv6 addresses, which map to 24.1k unique IPv4 prefixes across 7.5k ASes and 12.4k IPv6 prefixes across 7.6k ASes, respectively.
We use the extracted IP, prefix, and AS information throughout this paper.

\subsection{AS to Organization Mapping Datasets}\label{subsec:as-to-org}

For all ASes of sibling prefixes, we identify their organization using the AS organization dataset from Chen \etal \cite{chen2023improving}.
The dataset also allows us to identify sibling ASes, \ie ASes maintained by the same organization.
Due to the recency of this dataset, we use CAIDA's AS to organization mapping dataset \cite{caida_as2org} for the analysis prior to October 2022, and the Chen \etal \cite{chen2023improving} dataset for the analysis from October 2022 onward.
We use the AS to organization datasets throughout our analysis to provide insights on the status of sibling prefixes for the same organization compared to the rest of the data. 
We provide further details on sibling prefixes from the same organization in \Cref{subsec:evolution_of_sibling_prefixes}.

\subsection{Hypergiants and CDN Datasets}
\label{subsec:hg_dataset}

We use the list of hypergiant (HG) networks provided by Böttger \etal \cite{bottger2018looking} and Gigis \etal \cite{hypergiants2018gigis}, and the list of content distribution networks (CDNs) \cite{cdnplanet} to classify sibling prefixes as HG, CDN, and non-CDN-HG.
Using the HG and CDN datasets, we analyze the similarity status for sibling prefixes of CDN-HG versus non-CDN-HG in \Cref{subsec:sp_in_cdn_and_hgs}.

\subsection{ASdb Datasets}
\label{subsec:ASdb_dataset}

To infer the business type of the origin ASes for sibling prefixes, we use the ASdb dataset \cite{ziv2021asdb}.
The dataset classifies ASes into one or more of 17 business categories, such as ``Computer and IT'', ``Government'', or ``Media''.
This dataset allows us to analyze the prevalence of different business types among sibling prefixes (see \Cref{subsec:sibling_prefixes_business_type}).

\subsection{RPKI Dataset}
\label{subsec:rpki_dataset}

We examine the validity of each pair of sibling prefixes in the Resource Public Key Infrastructure (RPKI).
For this purpose, we download the RPKI data of all five Regional Internet Registries (RIRs) \cite{rpkidata} from September 2020 to September 2024 for every month.
With the RPKI data, we identify the route origin validation (ROV) state for sibling prefix pairs.
We use the RPKI dataset in \Cref{subsec:DS_prefix_rov}.

\subsection{Port Scan Dataset}
\label{subsec:port_scan_dataset}
We scan a set of 14 well-known ports \cite{bano2018scanlive,zakir2024tenyzmap,saltter2023brim} on all IP addresses of sibling prefixes. 
We report their responsive status and Jaccard similarity by comparing them to the Jaccard similarity observed in the OpenINTEL data \Cref{subsec:port_scan}. 
\section{Methodology}
\label{sec:method}

\begin{figure*}[!t]
      \centering
    \includegraphics[width=0.98\linewidth]{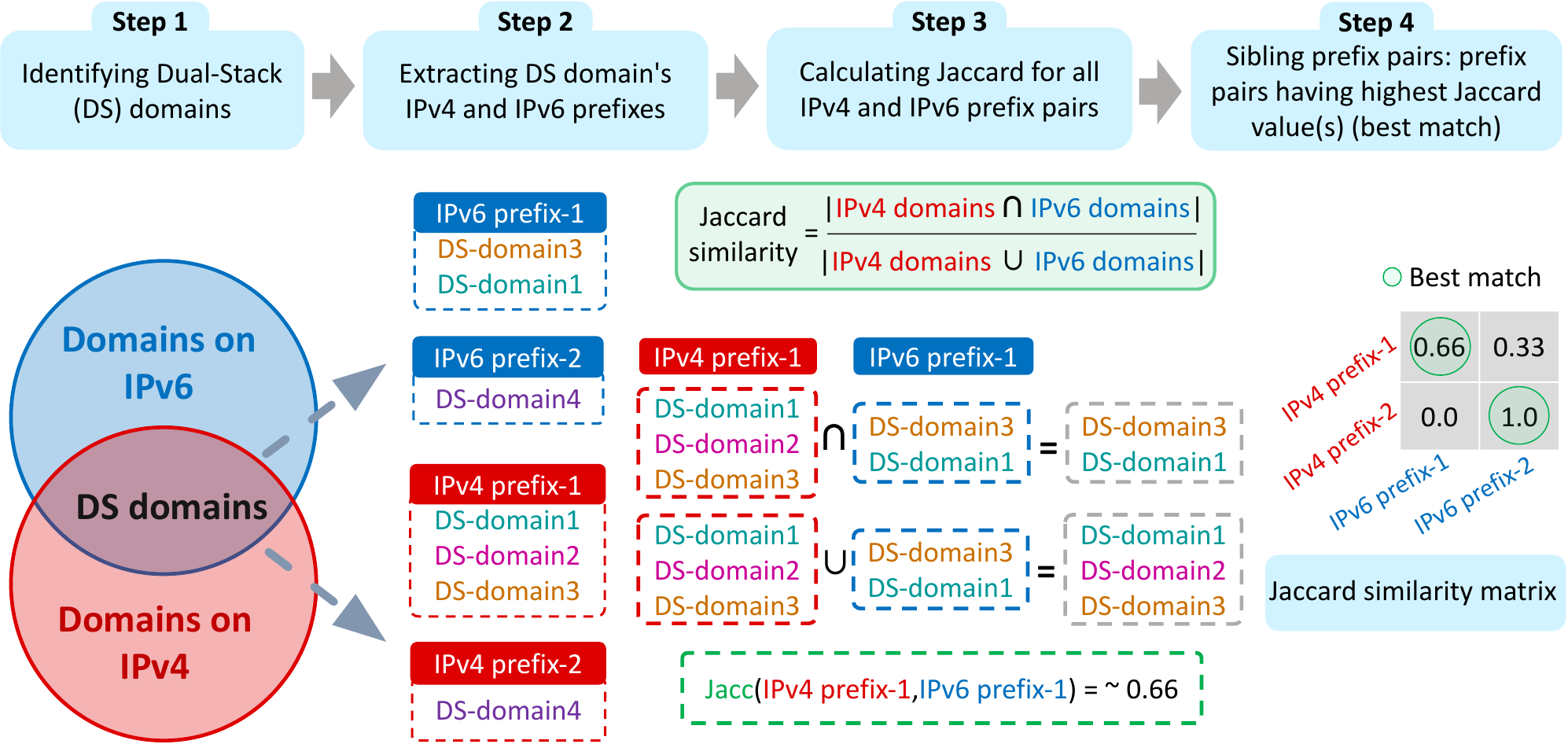}
    \caption{Methodology to identify sibling prefixes using Jaccard similarity index.} \label{fig:methodology_diagram_v2}
  \end{figure*}

This section explains our methodology to identify sibling prefixes.
We use the mapping of a domain name to an IPv4 and IPv6 address as an indicator for running services, which can then be used to infer sibling prefixes.

As we are interested in the actual domain that maps to an IP address, we use the domain name provided in the DNS response instead of the queried domain.
This helps to overcome cases where the domain name in the response is different from the domain that was queried, especially for DNS queries resulting in CNAME responses.
\Cref{fig:methodology_diagram_v2} provides an overview of our methodology to identify sibling prefixes.

\subsection{Identifying Sibling Prefixes}
\label{subsec:identifying-sibling-prefixes}

As shown in \Cref{fig:methodology_diagram_v2}, identifying sibling prefixes consists of four steps:
(1) Identifying dual-stack (DS) domains, (2) extracting DS domain's IPv4 and IPv6 prefixes, (3) calculating similarity metrics for all prefix pairs, and (4) selecting sibling prefixes.
We detail each of these steps as follow.

\parx{Step 1: Identifying dual-stack (DS) domains.}
In the first step, we examine the mapped IP addresses of all domain names in our dataset.
If the domain name maps to a CNAME, we follow the CNAME chain until we reach the final IP address in the CNAME chain.%
This results in two different sets of domains, all domains resolving to IPv4 addresses and all domains resolving to IPv6 addresses, as illustrated with red and blue circles in \Cref{fig:methodology_diagram_v2}.
The intersection of domain names in these two sets are domain names with both IPv4 and IPv6 addresses, \ie DS domains.
We use only these DS domains in our sibling prefix detection technique.
\parx{Step 2: Extracting DS domain's IPv4 and IPv6 prefixes.}
In the second step, we identify the prefixes for all IP addresses of DS domains.
We use the prefix information present in the OpenINTEL dataset (see \Cref{sec:dataset}) for every IP address of each DS domain.
We then group the domains by IPv4 and IPv6 prefixes to calculate the similarity between an IPv4 and an IPv6 prefix based on the set of DS domains.
\parx{Step 3: Calculating Jaccard value for all IPv4-IPv6 prefix pairs.}
In the third step, we calculate the Jaccard similarity index \cite{jaccard} for IPv4-IPv6 prefix pairs by comparing the set of domains resolving to each pair.
We denote all DS domains of an IPv4 prefix as set \(A\) and all DS domains of an IPv6 prefix as set \(B\).
Then, the Jaccard similarity index for sets \(A\) and \(B\) can be calculated as \Cref{eq:jacc}:

\begin{equation}\label{eq:jacc}
    \text{Jaccard}(A, B) = \frac{|A \cap B|}{|A \cup B|}\tag{1}
\end{equation}

Note that we also investigate alternative similarity metrics, see \Cref{subsec:other_similarity_methods}.

\parx{Step 4: Sibling prefix pairs.}
For each prefix pair, we get a Jaccard similarity value between 0 and 1, with values close to 1 indicating high similarity between two prefixes and values close to 0 indicating low similarity.
In the final step, we select the prefix pairs with the highest Jaccard value, \ie the ``best match''.
Prefix pairs with a similarity value of 0 are discarded, and if multiple prefix pairs share the same highest Jaccard value, we keep all of them.
We refer to these best matches as sibling prefix pairs for the corresponding IPv4 and IPv6 prefixes.
These are the sibling prefixes we analyze in-depth in the remainder of the paper.

\subsection{Examining Possible Similarity Metrics}
\label{subsec:other_similarity_methods}

Selecting a suitable similarity metric is crucial for identifying sibling prefixes.
We compare the overlap coefficient \cite{overlap_coefficient} and Dice coefficient\cite{dice} with the Jaccard similarity index \cite{jaccard}.

The overlap coefficient measures the overlap between two finite sets.
For two sets A and B, the overlap coefficient can be calculated using the intersection of sets A and B divided by the smaller size of the two sets.
To calculate the overlap coefficient (\(OC\)) for sets \(A\) and \(B\) we use \Cref{eq:oc}:

\begin{equation}\label{eq:oc}
    OC(A, B) = \frac{|A \cap B|}{\min(|A|, |B|)} \tag{2}
\end{equation}

The Dice coefficient is used to measure the similarity of two samples.
It is often used to compare the similarity of two texts or sequences of words.
The Dice coefficient for sets \(A\) and \(B\) can be calculated as shown in \Cref{eq:dice}:

\begin{equation}\label{eq:dice}
    \text{Dice}(A, B) = \frac{2 \times |A \cap B|}{|A| + |B|} \tag{3}
\end{equation}

Next, we identify sibling prefixes by applying the Jaccard similarity index, overlap coefficient, and Dice coefficient on DS domains and compare the results.
\Cref{fig:Jaccard_vs_Dice_vs_overlap_0day} shows the results of similarity values for the three metrics.
With the overlap coefficient more than 90\% of sibling prefix pairs have a similarity value of 1.0.
If either DS domain of the IPv4 prefix is a subset of the IPv6 prefix, or vice versa, the overlap coefficient value will be by definition 1.
This property makes the overlap coefficient unsuitable for our study; as explained, we are interested in similar prefixes, not overlapping prefixes.
The Dice and Jaccard lines in \Cref{fig:Jaccard_vs_Dice_vs_overlap_0day} are relatively similar, having a similarity value of 1 for around 50\% of sibling prefix pairs.
However, the similarity values below 1 have slightly different similarity values, with Dice being lenient to the right side and having higher similarity values.
This is because the Dice coefficient is generally more sensitive to slight overlaps in sets \cite{dice}.
On the other hand, the Jaccard index provides less biased and more balanced similarity results for variable set sizes \cite{fletcher2018comparing}.
In our study, as IPv4 and IP6 prefixes tend to have differently sized domain sets (\eg sometimes a couple of domains in one set compared to dozens in the other set), we choose the Jaccard index as a suitable similarity metric to identify sibling prefixes.

\begin{figure*}[!t]
  \minipage[t]{0.32\textwidth}
    \centering
    \includegraphics[width=\linewidth]{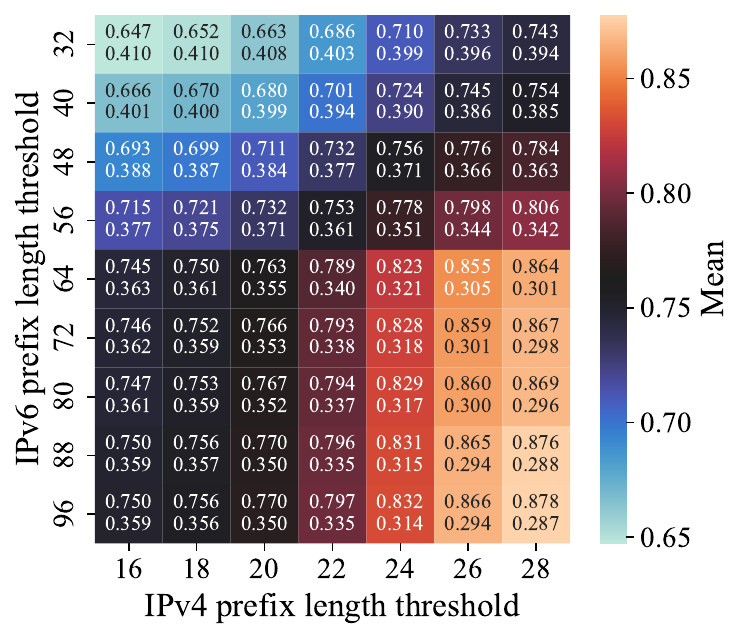}
    \caption{Heatmap of SP-Tuner Algorithm Performance: Mean Jaccard Index (top) and Standard Deviation (bottom) across IPv4 (x-axis) and IPv6 (y-axis) CIDR size thresholds.}
    \label{fig:sps_mean_std}
  \endminipage
  \hfill
  \minipage[t]{0.27\textwidth}
    \centering
    \includegraphics[width=\linewidth]{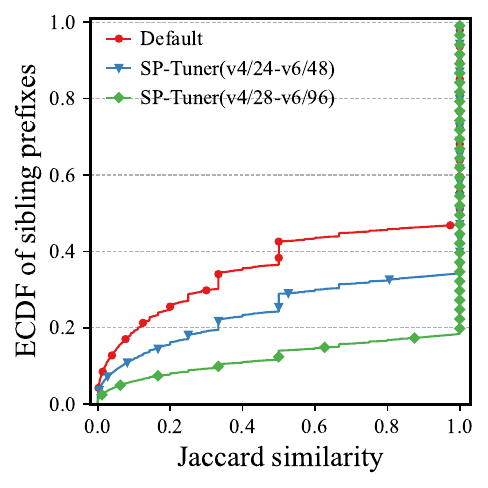}
    \caption{CDF of Jaccard similarity for sibling prefixes comparing default values with SP-Tuner algorithm at routable and optimal IPv4/IPv6 thresholds.}
    \label{fig:sp_tuner_ms}
  \endminipage
  \hfill
  \minipage[t]{0.34\textwidth}
  \includegraphics[width=\linewidth]{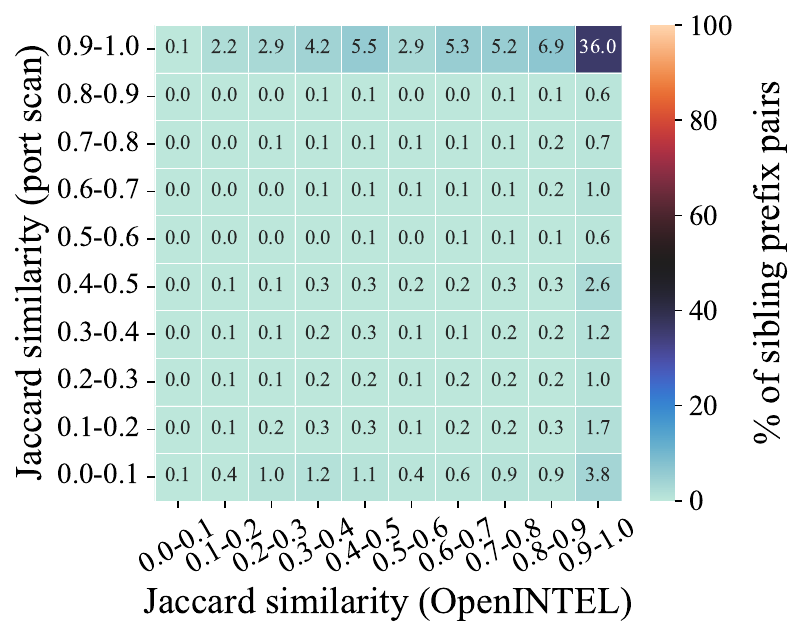}
  \caption{Heatmap of open port scanning: Each cell shows the percentage of sibling prefixes, with the x-axes and y-axes being the Jaccard values for DNS (OpenINTEL) and port scan data, respectively.}
  \label{fig:port_scan_perc_28_96}
\endminipage
\end{figure*}

    \subsection{Sibling Prefix Tuner (SP-Tuner) Algorithm}
    \label{sec:sp_tuner}

    We use IP prefixes from BGP announcements and domains to identify sibling prefixes. The CIDR sizes of these sibling prefix pairs, derived from BGP announcements, can occasionally be too specific. %
    Conversely, achieving better Jaccard similarity may be possible by changing the CIDR size of prefixes to cover more similar sets of domains.
    We introduce the sibling prefix tuner (SP-Tuner) algorithm to improve the similarity (\ie Jaccard similarity value) and fine tune the CIDR size of sibling prefixes.

    We investigate the effects of reducing and increasing prefix sizes on Jaccard similarity of sibling prefixes. 
    Two possible approaches are to either check for the Jaccard similarity of less specific or more specific of the current sibling prefix pairs.
    For example, to examine the covering /22 prefix of a BGP-announced /23, or to examine the Jaccard similarity for more specific prefixes, like the more specific /24 prefix of the BGP-announced /23 prefix.
    We implement and thoroughly analyze both the less specific (see \Cref{subsec:sp_tuner_ls} for more details) and more specific variants of SP-Tuner algorithm. 
    In the remainder of the paper we leverage SP-Tuner using more specific CIDR sizes as it produces better Jaccard similarity values.
    A more specific CIDR block, like /28 instead of /24, gives network operators tighter control over which IP addresses are affected by a policy.
    This level of precision is beneficial when network operators want to block a small set of suspicious IPs without disrupting legitimate traffic or when allowing access only to a narrow group of trusted hosts.
    Broader prefixes can be too blunt and end up causing unintended side effects.
    We implement the SP-Tuner algorithm with two PyTricia tree \cite{pytricia} data structures for each IP version and their respective DS domains.
    PyTricia facilitates efficient storage and retrieval of IP addresses and their associated domains within a tree data structure.
    
    As shown in \Cref{alg:sp_tuner_ms}, the more-specific variant of SP-Tuner (\ie SP-Tuner-MS) processes each sibling prefix pair to refine it into more specific subprefixes.
    The algorithm traverses each tree in the downward direction from the point of prefix insertion to identify more specific prefixes with improved Jaccard values.
    During this traversal, some domains may fall under alternate branches that are not part of the best-matching subprefix. These subprefixes still contain relevant domains but are excluded from the maximal pair.
    To prevent domain loss, we track these additional branches and treat them as new candidate sibling prefix pairs, applying the SP-Tuner algorithm to them as well.
    This approach ensures that no domains are lost during prefix size tuning. 
    Additionally, we employ thresholds at both IPv4 and IPv6 prefix levels to ensure that tree traversal stops at meaningful levels.

    \begin{figure}[t]
      \begin{minipage}{0.45\textwidth}
        \begin{algorithm}[H]
          \caption{SP-Tuner-MS (More Specific)}\label{alg:sp_tuner_ms}
          \scriptsize

          \begin{algorithmic}[1]
            \Statex \textbf{Data:} DS domains, IPv4\_addresses, IPv6\_addresses, sibling prefix pairs, Jaccard similarity
            \Statex \textbf{Result:} Refined sibling prefix pairs with improved Jaccard similarity
            \Statex \textbf{Initialization:}
            \Statex \hspace{\algorithmicindent} $\textit{tree\_v4} \gets \{\textit{DS\_domains}, \textit{IPv4\_addresses}\}$
            \Statex \hspace{\algorithmicindent} $\textit{tree\_v6} \gets \{\textit{DS\_domains}, \textit{IPv6\_addresses}\}$
            \Statex \hspace{\algorithmicindent} $\textit{curr\_jacc} \gets \textit{precomputed non-zero value}$ %
            \Statex \hspace{\algorithmicindent} $\textit{prefix\_v4\_len\_thresh}$ \Comment{IPv4 prefix length threshold}
            \Statex \hspace{\algorithmicindent} $\textit{prefix\_v6\_len\_thresh}$ \Comment{IPv6 prefix length threshold}

            \For{$(\textit{sibling\_prefix\_v4,\ sibling\_prefix\_v6}) \in \textit{sibling\_prefix\_pairs}$} 
              \State $\textit{tree\_v4} \gets \textit{sibling\_prefix\_v4}$
              \State $\textit{tree\_v6} \gets \textit{sibling\_prefix\_v6}$
              \State $\textit{new\_jacc} \gets 0$
              \While{$\textit{new\_jacc} \leq \textit{curr\_jacc}$}
                  \State $\textit{prefix\_v4\_subprefixes} \gets \textit{GetNextSubprefixes}(\textit{sibling\_prefix\_v4})$
                  \State $\textit{prefix\_v6\_subprefixes} \gets \textit{GetNextSubprefixes}(\textit{sibling\_prefix\_v6})$
                  \For{$\textit{prefix\_v4} \in \textit{prefix\_v4\_subprefixes}$}
                      \For{$\textit{prefix\_v6} \in \textit{prefix\_v6\_subprefixes}$}
                          \State $\textit{new\_jacc} \gets \textit{max}(\textit{Jaccard}(\textit{prefix\_v4}, \textit{prefixv6}))$
                          \If{$\textit{HasBranch}(\textit{prefix\_v4}, \textit{prefix\_v6})$}
                              \State $\textit{sibling\_prefix\_pairs} \gets \textit{UpdateBranches()}$
                          \EndIf
                      \EndFor
                  \EndFor
                  \If{$(\textit{prefix\_v4\_len} \geq \textit{prefix\_v4\_len\_thresh})$ \\
                    \hspace{2em} \textbf{and} $(\textit{prefix\_v6\_len} \geq \textit{prefix\_v6\_len\_thresh})$}
                      \State \Return $(\textit{prefix\_v4}, \textit{prefix\_v6})$ %
                  \EndIf
              \EndWhile
            \EndFor
          \end{algorithmic}
        \end{algorithm}
      \end{minipage}
    \end{figure}

  \subsection{Effectiveness of SP-Tuner}
  \label{subsec:sp-tuner_impact}

  To evaluate the effectiveness of the SP-Tuner algorithm, we apply the algorithm on the most recent domain snapshot from Septeber 11, 2024.
  We configure two CIDR size values, one for IPv4 and one for IPv6, as the threshold of the SP-Tuner algorithm. 
  The SP-Tuner algorithm calculates the Jaccard value of sibling prefixes by reducing the sibling prefix size up to and including the threshold to check and improve the Jaccard value whenever possible. 
  To assess the impact of different prefix size thresholds, we investigate various IPv4 and IPv6 prefix sizes, starting from /16 up to /31 for IPv4, and from /32 up to /124 for IPv6 sibling prefixes, respectively.

  \Cref{fig:sps_mean_std} shows our results for SP-Tuner thresholds up to /28 IPv4 and /96 IPv6 prefix sizes.
  We refer the interested reader to \Cref{fig:sp_tuner_mean_std_all} in \Cref{subsec:sp_tuner_sensitivity} for the complete heatmap.
  The color in \Cref{fig:sps_mean_std} and the top value within each cell indicate the mean Jaccard value for all sibling prefixes, and the bottom value within each cell shows the standard deviation of the Jaccard values. 
  Moreover, the x-axis and y-axis in \Cref{fig:sps_mean_std} represents IPv4 and IPv6 prefix length thresholds, respectively, starting from /16 for IPv4 and /32 for IPv6, with subsequent thresholds increasing incrementally.

  Examining the mean Jaccard values of \Cref{fig:sps_mean_std} row-wise for the IPv4 threshold shows that the more specific the IPv4 CIDR size, the higher the mean Jaccard value.
  The same pattern applies to the column-wise mean Jaccard values for the IPv6 threshold.
  We find that from the lowest mean Jaccard value of 0.647 for the combination IPv4 threshold of /16 and IPv6 threshold of /32, the SP-Tuner algorithm can improve the mean Jaccard value up to 0.878 for the /28 and /96 threshold values of IPv4 and IPv6 prefixes, respectively.
  On the same combination, SP-Tuner also reduces the standard deviation from 0.410 to 0.287.
  The improvements in Jaccard similarity, deriving more similar sibling prefixes for existing prefixes with low standard deviation, show that the SP-Tuner algorithm is able to identify more fine-tuned sibling prefixes with higher Jaccard similarity.
  Note that the goal is to identify sibling prefixes at the prefix level, not the IP address level.
  Hence, we pick prefix thresholds of /28 and /96.
  We leave the choice to users of sibling prefixes to pick suitable CIDR sizes based on their specific use case, be it the default BGP-announced sibling prefixes as seen in the dataset, or /24 IPv4 and /48 IPv6 thresholds for most-specific routable prefixes, or /28 and /96 as in our case.
  For this study, we provide a detailed analysis based on the /28 and /96 threshold prefixes as the highest Jaccard similarity value with the lowest standard deviation.
  
  We further examine the impact of the SP-Tuner algorithm by applying it on the data from September 11, 2024 and comparing it with the default BGP-announced, as seen in the DNS data, sibling prefixes. 
  \Cref{fig:sp_tuner_ms} shows the CDF of Jaccard similarity values for the default BGP-announced sibling prefixes in red, SP-Tuner routable threshold of /24 and /48 for IPv4 and IPv6, respectively, and the best performance SP-Tuner case with an IPv4 threshold of /28 and an IPv6 one of /96.
  We see for the default case around 52\% of sibling prefixes have a Jaccard value of 1, \ie they are perfect matches.
  After applying SP-Tuner, this percentage increases to 67\% for the routable prefix threshold and to more than 82\% for the /28 and /96 thresholds.
  In summary, the SP-Tuner with the best performance thresholds increases the number of sibling prefixes from 52\% in the default case to around 82\% of sibling prefix pairs, all having Jaccard similarity index of 1 and more tailored smaller CIDR sizes by excluding irrelevant address spaces of prefixes.

  \subsection{Ground Truth Evaluation}\label{subsec:ground_truth}

  To compare our identified sibling prefixes with a real-world deployment, we use RIPE Atlas probes \cite{ripeatlas_probes} as a ``ground-truth'' dataset.
  We extract the publicly available IPv4 and IPv6 addresses from dual-stack RIPE Atlas probes, map them to prefixes, and compare these prefixes to our identified sibling prefixes.
  
  From a total of 5174 dual-stack probes, 2200 (42.5\%) probes have IPv4 and IPv6 covered by our sibling prefixes, 1663 (32.1\%) of probes are partially covered by our sibling prefixes (\ie either IPv4 or IPv6), and 1310 (25.3 \%) probes are not covered by our sibling prefixes at all.
  This shows that almost half of the dual-stack probes are completely covered by our sibling prefix dataset, which means that we can evaluate the similarity of the RIPE Atlas dual-stack probes with our sibling prefix results. Out of the 2200 completely covered RIPE Atlas probes, 1966 (89.36\%) are in our best-match sibling prefixes (\ie results of step 4 of \Cref{fig:methodology_diagram_v2}).
  On the other hand, only 234 (10.64\%) of the completely covered probes are not best-match siblings.
  This underlines that our sibling prefixes line up quite well with the RIPE Atlas probes ground-truth dataset, which increases the confidence in our chosen methodology.

  Finally, we also check our identified sibling prefixes against 260 globally distributed dual-stacked virtual private servers (VPSes) from IPinfo's probe network \cite{probenet} hosted by different VPS providers (\eg Google, Azure, Vultr, AWS).
  For the VPSes where we get both an IPv4 and IPv6 address match in our siblings, 53 are within our best-match siblings, whereas 13 are mismatches, with the remainder not having IPv4 and IPv6 matches.
  This again shows, that the majority of our best matches provide accurate siblings matches.

  \begin{figure*}[!t]
    \centering
    \includegraphics[width=\linewidth]{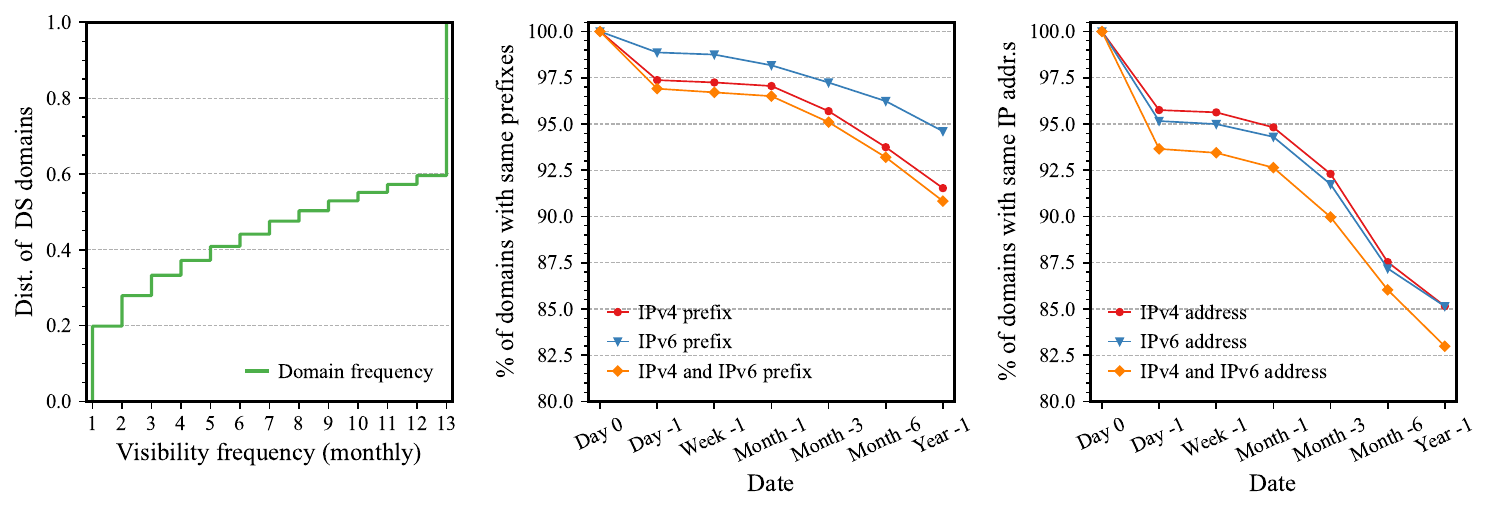}
    \caption{Consistently visible DS domains (left), prefix changes (center), and address changes (right).}
    \label{fig:domain_stability_analysis}
  \end{figure*}

\subsection{Overlap with Open Ports}
\label{subsec:port_scan}

To better understand if our identified sibling prefixes based on domain names also align with actual deployed infrastructure, we compare our domain-based approach with an open-port-based approach.
We use IP addresses from our latest OpenINTEL dataset on September 11, 2024 as the reference dataset and perform active scanning on September 19, 2024.
We then perform port scans using ZMap \cite{zmap} and ZMapv6 \cite{zmapv6} on 14 common open ports \cite{bano2018scanlive,zakir2024tenyzmap,saltter2023brim}, \ie 20, 21, 22, 23, 25, 53, 80, 110, 123, 143, 161, 194, 443, 7547.
For 70.9\% of sibling prefixes we get responses. 
Then, we map the addresses from the port scanning results with the prefixes in sibling prefixes to identify the associated prefixes for them.
We use the Jaccard similarity index to determine the similarity between these prefix pairs.
However, instead of using the domain set, we utilize the responsive ports within IPv4 and IPv6 prefixes. 

In \Cref{fig:port_scan_perc_28_96} the heatmaps display the Jaccard similarity values of sibling prefixes based on the port scan on the y-axis, and the Jaccard values for the same pairs of sibling prefixes based on the OpenINTEL dataset on the x-axis.
The color and number in each heatmap cell shows the percentage of sibling prefixes based on 28-96 threshold pair of SP-Tuner algorithm.
For the top row, the rightmost cell shows a value of 36\% as the highest number of sibling prefixes matching $\geq0.9$ Jaccard value on the both x and y axes. This indicates a correlation between the similarity of the OpenINTEL results and port scan results for sibling prefixes.
Moreover, the values in the topmost row of \Cref{fig:port_scan_perc_28_96} reflect that for sibling prefix pairs having high Jaccard value ($\geq0.9$) in port scan, the OpenINTEL Jaccard similarity is distributed to a lower range.
A similar pattern with a slightly lower percentage is valid by comparing the OpenINTEL Jaccard value with $\geq0.9$ on the rightmost column to the port scan on the y-axis.
In summary, 36\% of responsive sibling prefix pairs with a high Jaccard similarity of $\geq0.9$ on OpenINTEL dataset are also likely to have a high Jaccard value based on port scan results.

\subsection{Limitations}\label{subsec:dataset_limitations}
Although we try to fine-tune our methodology quite well to the task of sibling prefix detection, there are some limitations to it, which we will elaborate in the following.
As our sibling prefix methodology is fundamentally based on domain name data, it is only as good as the used domain name data for sibling identification.
In our analysis, we leverage a large-scale DNS resolution dataset (see \Cref{subsec:dns_dataset}).
Even though the dataset covers more than 13 million unique domains, our findings are limited to the used dataset, which represents only a subset of all DNS domains. %
Thus, adding even more domain name data as an input to the methodology, might provide better coverage in terms of the number of sibling prefixes and potentially also their quality.
Our methodology can also be applied with inputs different than domain names, such as alias datasets or open ports on devices (see \Cref{subsec:port_scan}).
As long as these inputs result in a mapping from a prefix to a set, our technique of using set similarity and then picking the maximum similarity value can still be applied.
Despite these limitations, we believe that our methodology is suitable to detect sibling prefixes in the Internet.

\subsection{Ethical Considerations}

In this work, we follow ethical Internet measurement guidelines \cite{zmap,dittrich2012menlo,partridge2016ethical}.
First, we use use publicly available datasets (see \Cref{sec:dataset}).
Second, for ground truth evaluation and comparison of DNS data with open ports (see \Cref{subsec:ground_truth,subsec:port_scan}), we perform port scans on IP addresses from the OpenINTEL dataset.
To limit potential harm for third parties, we use a blocklist, maintain public web pages describing our measurement activity and providing contact information, and we limit our scanning rate to 50 kpps.
Third, during our measurements, we did not receive any complaints.
Thus, we believe that our work does not raise ethical concerns.
 
  \begin{figure*}[!t]
    \minipage[t]{0.36\textwidth}
      \centering
      \includegraphics[width=\linewidth]{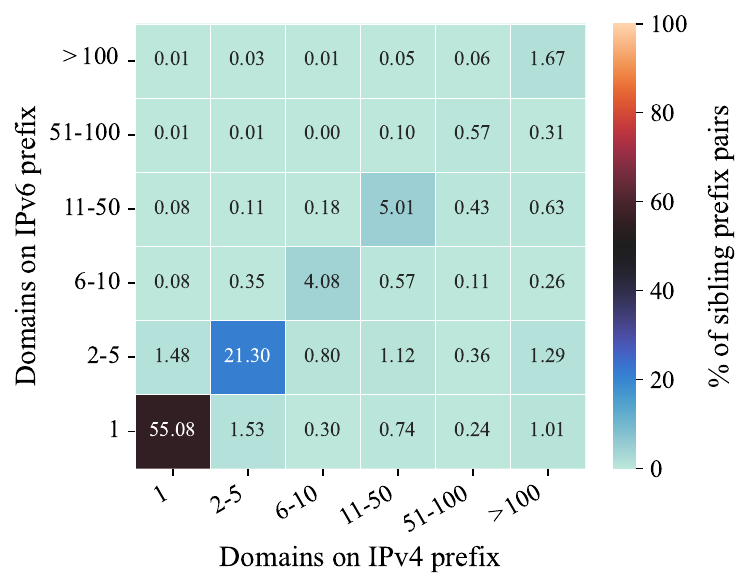}
      \caption{Sibling prefix pairs count and classification based on number of dual-stack domains in each IPv4 and IPv6 prefix of sibling prefixes}
      \label{fig:sp_domain_category_28_98}
    \endminipage
    \hfill
    \minipage[t]{0.28\textwidth} 
      \centering
      \includegraphics[width=\linewidth]{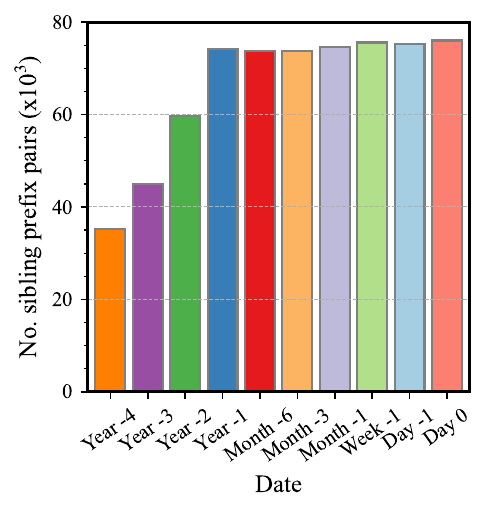}
      \caption{Number of sibling prefixes at different points in time. (Reference date: September 11, 2024)}
      \label{fig:Jaccard_barplot_28_96}
    \endminipage
    \hfill
    \minipage[t]{0.29\textwidth}
      \centering
      \includegraphics[width=\linewidth]{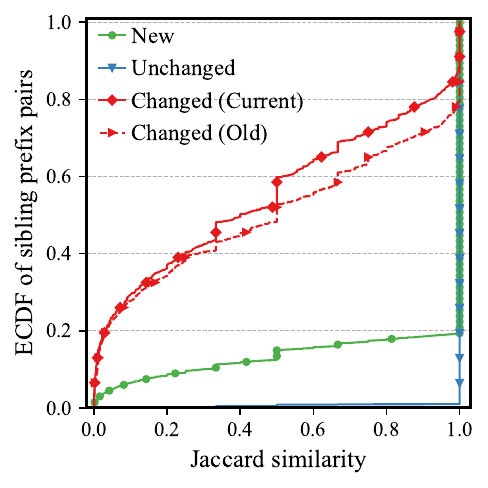}
      \caption{Improvement of Jaccard similarity of sibling prefixes over time.}
      \label{fig:Jaccard_improvement_over_time_28_96}
    \endminipage
  \end{figure*}

\section{Sibling Prefixes}
\label{sec:sibling_prefixes}

In this section, we thoroughly analyze sibling prefixes observed in our dataset. We begin with examining IP and prefix changes of dual-stack domains, followed by exploring their distribution in sibling prefixes.
We then continue exploring sibling prefixes by conducting a longitudinal analysis and by assessing the impact of SP-Tuner on historical data.
Next we analyze CIDR sizes, origin ASes, and business types of sibling prefixes.
Finally, we conclude the analysis by investigating differences of sibling prefix characteristics in hypergiant and CDN networks and examine their ROV status.

Throughout this section, we use the sibling prefix pairs resulting from our SP-Tuner algorithm with a /28 IPv4 and a/96 IPv6 prefix threshold.
For some cases where using sibling prefixes as observed in BGP announcements (\ie the default case) is more relevant to our analysis, we use that notion of sibling prefixes.
Analyzing the OpenINTEL dataset, we identify 3.5M dual-stack (DS) domains running on IP addresses of 46.3k IPv4 and 39.5k IPv6 prefixes. These prefixes originated from  6.6k IPv4 and 5.9k IPv6 ASes, collectively they result in 76k sibling prefixes as of September 2024.

\subsection{Address and Prefix Dynamics in Dual-Stack Domains}
\label{subsubsec:ds_prefix_dynamics}

Before diving into the analysis of sibling prefixes themselves, we first analyze address and prefix dynamics found in DS domains.
The goal of this analysis is to understand to what extent addresses and prefixes of DS domains change over time.
We analyze address and prefix changes for the DS that are consistently visible in our dataset.
To identify these consistent DS domains, we examine how frequently a sibling prefix appears in our dataset over a one year period from September 13, 2023, to September 11, 2024, \ie a total of thirteen monthly snapshots.
We collect data over a 13-month period to ensure year-over-year coverage and make full use of available data.

The left subplot in \Cref{fig:domain_stability_analysis} illustrates the cumulative distribution of DS domains on the y-axis and the number of times a DS domain appeared in the data (``visibility frequency'') on the x-axis.
Around 40\% of DS domains are consistently visible in all thirteen data points. About 20\% of DS domains appear only once, and the remaining 40\% of DS domains have a visibility frequency of two to twelve snapshots.
These percentage values indicate that apart from the 40\% consistent DS domains, the remaining 60\% of DS domains are either not consistently dual-stack (\ie they switch between dual-stack to IPv4/IPv6-only) or are not always present in our dataset.

Next, we want to analyze and peek into the temporal prefix stability of the 40\% consistent DS domains.
The center subplot of \Cref{fig:domain_stability_analysis} shows the percentage of DS domains having stable prefixes over time, again with the reference date of September 11, 2024, as ``day 0'' on the x-axis.
We observe that around 95\% of consistently visible DS domains are resolved to the same prefix for up to three months.
We do not see any prefix change throughout the one-year analysis period for more than 91\% of DS domains.
While the maximum change for IPv4 prefixes is around 9\%, this number is lower at around 6\% for IPv6 prefixes.
This shows that overall, consistent DS domains exhibit high prefix stability.

Finally, we analyze changes of consistently visible DS domains on the IP address level, shown in the right subplot of \Cref{fig:domain_stability_analysis}.
We find that 83\% of DS domains are mapped to the same IP addresses throughout the analysis period.
Overall, on the IP address level IPv4 and IPv6 behave much more similarly compared to the prefix level.
This indicates that addresses change at a similar rate in both protocols, but prefixes are more stable in IPv6 compared to IPv4.

\textbf{To summarize:} We show that 40\% of DS domains are consistently visible for thirteen months.
Additionally, 83\% of these consistent DS domains have stable addresses, and 91\% of the consistent DS domains have stable prefixes throughout one year.
This shows that consistent DS domains from our DNS dataset are stable and can therefore be used for our sibling prefix identification technique.

\subsection{Dual-Stack Domains in Sibling Prefixes}
\label{subsec:domains_in_sibling_prefixes}
  
\begin{figure*}[!t]
    \minipage[t]{0.28\textwidth}
      \centering
      \includegraphics[width=\linewidth]{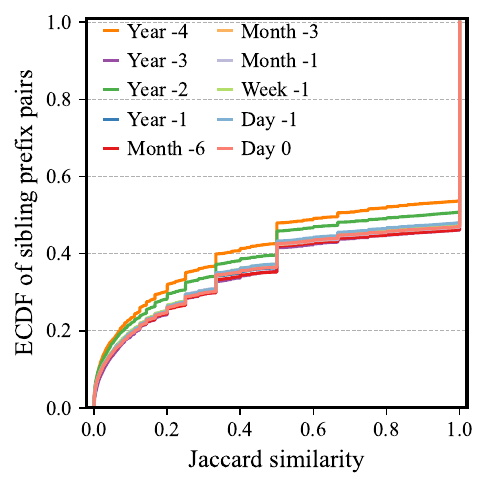}
      \caption{Jaccard similarity of sibling prefixes for various data points in time. (Reference date: September 11, 2024)}
      \label{fig:Jaccard_over_time_analysis_default}
    \endminipage
    \hfill
    \minipage[t]{0.28\textwidth}
    \centering
    \includegraphics[width=\linewidth]{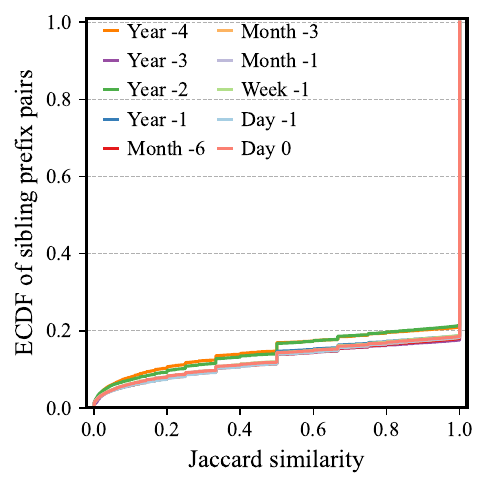}
    \caption{CDF of Jaccard similarity values for sibling prefixes on various data snapshots in time. }
    \label{fig:Jaccard_over_time_analysis_28_96}
    \endminipage
      \hfill
    \minipage[t]{0.36\textwidth}
    \centering
    \includegraphics[width=\linewidth]{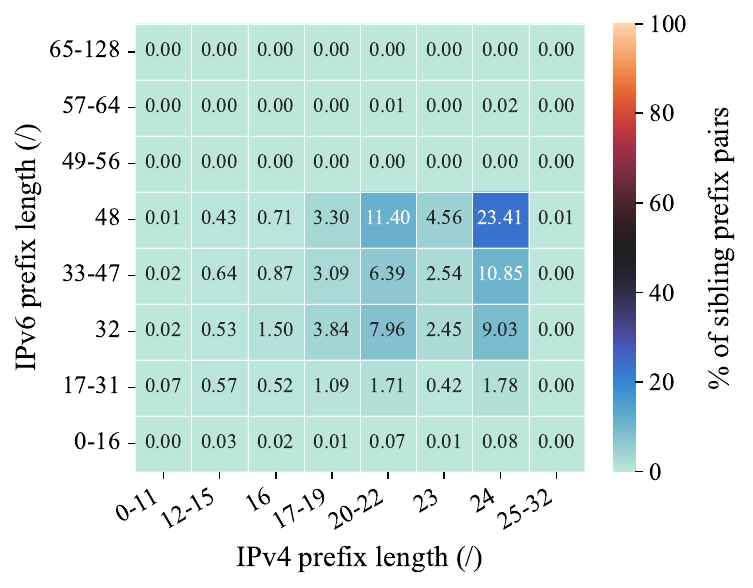}
    \caption{Distribution of CIDR sizes in sibling prefixes.}
    \label{fig:sp_prefix_size}
    \endminipage
  \end{figure*}

Next, we focus on analyzing the number of DS domains within each sibling prefix pair.
For this purpose, we first generate sibling prefixes using the SP-Tuner algorithm with a /28 and /96 CIDR threshold. Then, we group sibling prefixes by the number of dual-stack domains within each pair.
As shown in \Cref{fig:sp_domain_category_28_98}, we bin sibling prefixes regarding the number of associated domains.
The cell color and number in the heatmap show the percentage of sibling prefixes for each group of DS domains.
Over 55\% of sibling prefix pairs contain a single domain, as shown in the bottom left corner of the plot. 
The sibling prefix pairs with 2 to 5 domains are the second-largest group, making up 21.3\% of sibling prefixes.
Furthermore, sibling prefixes with more than 100 domains for IPv4 and IPv6 contribute 1.6\% to all sibling prefixes.
Finally, the diagonal cells in the heatmap contains a relatively high percentage of sibling prefixes compared to neighboring cells, indicating that sibling prefixes are relatively likely to share a similar number of DS domains for IPv4 and IPv6 prefixes.

\textbf{To summarize:} The majority of sibling prefixes contain single domains.
Moreover, we observe a relatively high number of IPv4 sibling prefixes with more than 100 dual-stack domains.
Finally, sibling prefixes tend to have a similar number of domains for IPv4 and IPv6.

\subsection{Longitudinal Analysis}
\label{subsec:longitudinal-analysis-of-sibling-prefixes}

In this part of the analysis we perform a longitudinal analysis of sibling prefixes over a 4-year period, as shown in \Cref{fig:Jaccard_barplot_28_96}.
The x-axis shows different points in time starting from the base date September 11, 2024 denoted as ``day 0'', and previous points in time denoted by the difference to this base date.
Over the past four years, the number of sibling prefixes has doubled from around 36k to more than 76k.
This is due to the addition of dual-stack domains in OpenINTEL (see \Cref{subsec:dns_dataset}) and to the increased use of IPv6 among previous IPv4-only domains.

\begin{figure*}[!t]

  \minipage[t]{0.31\textwidth}
    \centering
    \includegraphics[width=\linewidth]{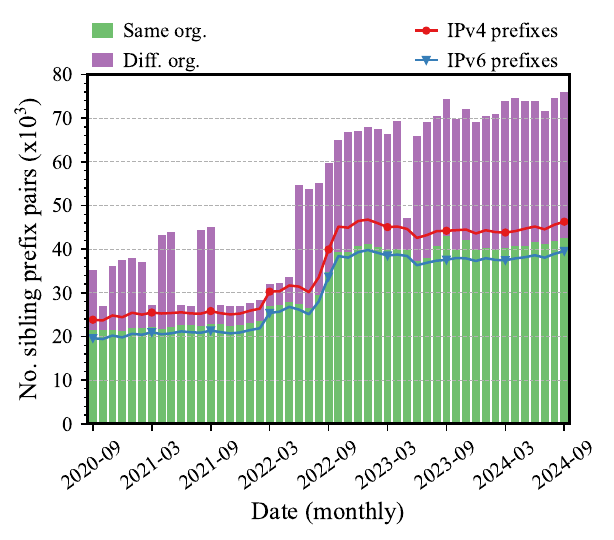}
    \caption{Sibling prefix pairs count for the same and different organizations over time. Lines represent the number of unique IPv4 and IPv6 prefixes in sibling prefixes.} 
  \label{fig:same_diff_orga_over_time}
  \endminipage
  \hfill
  \centering
  \begin{minipage}[t]{0.31\textwidth}
      \centering
      \includegraphics[width=\linewidth]{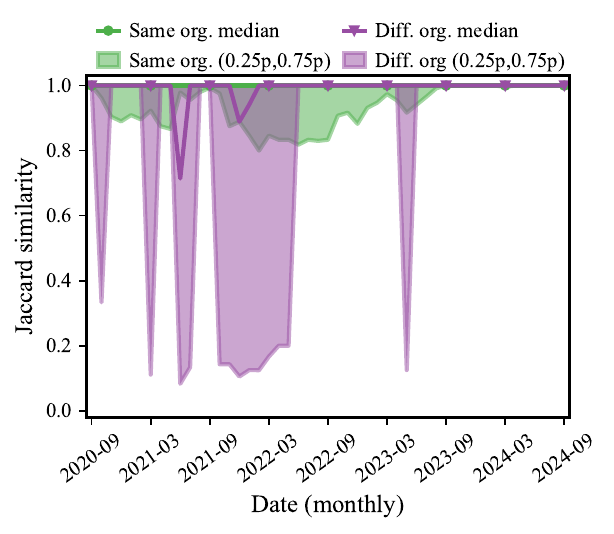}
      \caption{Median Jaccard similarity values of sibling prefix pairs for same and different organizations over time.}
      \label{fig:jacc_over_time_same_diff_orga_28_96}
  \end{minipage}
  \hfill
  \minipage[t]{0.33\textwidth}
  \centering
  \includegraphics[width=\linewidth]{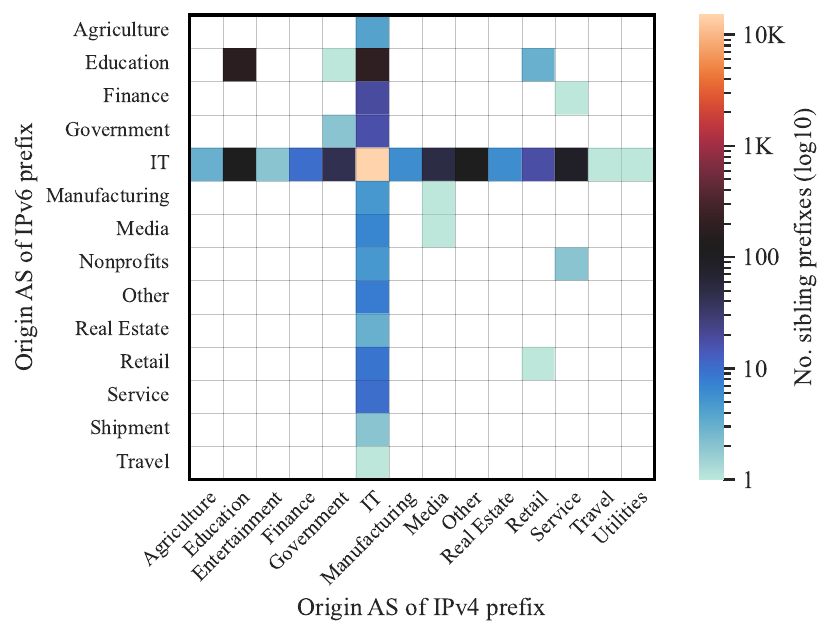}
  \caption{Business types heatmap for origin ASes of sibling prefixes, with IPv4 on x-axis and IPv6 on y-axis, indicating the count of sibling pairs in each cell.}%
    \label{fig:sibling_prefixes_in_ASdb}
    \endminipage
\end{figure*}

To provide insights into temporal effect on sibling prefixes, we perform a longitudinal analysis of Jaccard values.
To this end, we compare our most recent snapshot (September 2024) with data from four years ago (September 2020).
Note that---as any longitudinal analysis---this analysis is generally affected by changes in the underlying used datasets as well as changes in the Internet as a whole.
When investigating the potential impact of changes in the datasets, we can see that the percentage of DS domains remains very stable, even for adding the ``.fr'' TLD in August 2022 and removing the Alexa toplist from the OpenINTEL data in May 2023 (see \Cref{fig:ds_overtime} right).
When looking at the number of sibling prefixes in \Cref{fig:same_diff_orga_over_time} we see a slight changes for these two dates.
Rather than sudden jumps or drops these changes are more gradual, hinting at multiple different effects being at play in addition to changes in the underlying datasets.
Therefore, we believe that the general trends in the longitudinal analaysis still provide valuable insights into sibling prefix changes over time.

\Cref{fig:Jaccard_improvement_over_time_28_96} compares the Jaccard value distribution for sibling prefixes where the Jaccard value has not changed (``unchanged'', blue line), with changed sibling prefixes (``changed'', red lines), and completely new sibling prefixes (``new'', green line).
We further differentiate the changed sibling prefixes by plotting their old Jaccard value (dashed line) and new Jaccard value (solid line).

From a total of 76k sibling prefixes, the ``new'' category makes up 67k (88\%), ``unchanged'' 7.7k (10\%), and ``changed'' 1.2k (2\%) of sibling prefixes.
We find that unchanged sibling prefixes exhibit the highest similarity, with almost all of them having a Jaccard value of 1.
New sibling prefixes are more similar (80\% with Jaccard value 1) compared to old, but changed (21\% with Jaccard value of 1) sibling prefixes.
Finally, the current Jaccard value of changed sibling prefixes is lower compared to their Jaccard value four years ago.

To measure the impact of SP-Tuner on Jaccard similarity, we compare sibling prefixes as observed in our dataset (\ie with default BGP-announced prefix sizes) in \Cref{fig:Jaccard_over_time_analysis_default} with sibling prefixes after applying SP-Tuner in \Cref{fig:Jaccard_over_time_analysis_28_96}.
For the default case, we see that around 45 to 55\% of sibling prefixes have a Jaccard similarity value of 1.
However, after applying SP-Tuner in \Cref{fig:Jaccard_over_time_analysis_28_96}, the percentage of sibling prefixes with a Jaccard similarity value of 1 is almost doubled to around 80\%.
This improvement in Jaccard similarity values shows that SP-Tuner is able to identify a high number of fine-tuned sibling prefixes for various points in time. %

\textbf{To summarize:} %
The number of sibling prefixes has more than doubled compared to four years ago to around 76k in September 2024.
We find that the Jaccard value of sibling prefixes has substantially improved over time, with old prefixes having around 21\% and new sibling prefixes having 80\% of prefix pairs with a Jaccard value of 1.
The changed sibling prefixes have the least, 18\%, and unchanged sibling prefixes have the highest percentage of 99\% sibling prefixes with a Jaccard value of 1.
Moreover, fine-tuned sibling prefixes identified by SP-Tuner outperform BGP-announced sibling prefixes by almost doubling the fraction of perfect matches.

\subsection{CIDR Sizes}
\label{subsec:sp_prefix_size}

We continue our analysis by examining the most common CIDR sizes of IPv4 and IPv6 in sibling prefix pairs. As for SP-Tuner, the vast majority, \ie 86.95\%, of sibling prefixes fall into /28 and /96 IPv4 and IPv6 CIDR sizes.
This is due to the /28–/96 threshold applied by SP-Tuner, which maps most prefixes to these prefix length values (see \Cref{subsec:sp_tuner_sensitivity}, \Cref{fig:sp_prefix_size_28_96_p}).
Therefore, we focus on the default case as it shows a more nuanced distribution as can be seen in \Cref{fig:sp_prefix_size}.
To aid readability, we group sibling prefixes into different CIDR size groups, separating out commonly used CIDR sizes.
We observe that /24 is the most prevalent CIDR size in IPv4, while /48 is most prominent in IPv6.
Consequently, the /24-/48-combination makes up the largest share of sibling prefixes with 23.41\% of pairs.
Notably, we see relatively high percentages for cells in the region between /17--/24 in IPv4 and /32--/48 in IPv6, which together makes up more than 88\% of all sibling prefixes.
We observe a small fraction of sibling prefix pairs formed by less specific prefix CIDR sizes smaller than /17 for IPv4 and less specific than /32 in IPv6.
Moreover, non-globally-routable prefixes \cite{sediqi2022hyper} more specific than /24 in IPv4 and /48 in IPv6 are very rare among sibling prefixes.

\textbf{To summarize:}
The majority of SP-Tuner sibling prefixes fall into the /28-/96 group, whereas BGP-announced sibling prefixes are most commonly /24 in IPv4 and /48 in IPv6.
Moreover, CIDR size ranges from /17 to /24 in IPv4 and /32 to /48 in IPv6, make up the vast majority of sibling prefixes on the Internet.

\subsection{Origin ASes}
\label{subsec:evolution_of_sibling_prefixes}

In this section, we analyze the difference of sibling prefixes being originated by the same or different organizations.
We merge sibling ASes in order to obtain sibling prefixes originated by the same or different organizations \cite{as2org}. 
If the origin ASes of IPv4 and IPv6 sibling prefixes in our dataset either share the same AS number or, if different, are registered to the same organization name, we classify them as ``same organization''; the remaining sibling prefixes are classified as ``different organization''.
\Cref{fig:same_diff_orga_over_time} shows a longitudinal view of the number of sibling prefixes originated by the same or different organization category.
We find that of the nearly 76k sibling prefixes in September 2024, more than half of them (around 41k) belong to the same organization category, indicating that their IPv4 and IPv6 origin ASes share the same organization name.
Note that this does not mean that all of these prefixes belong to a single organization, but rather to multiple organizations whose IPv4 and IPv6 ASes are registered under the same name.
Sibling prefix pairs from the different organizations category contribute around 35k pairs in September 2024.
We also notice a dip in sibling prefixes originated by different organization ASes in May 2023 and earlier snapshots.
Investigating the case, we find a unique domain (\textit{site24x7.enduserexp.com}) missing in the OpenINTEL data on May 2023 and in earlier snapshots.
Site24x7 is an IT infrastructure monitoring company \cite{site24x7}, hosting probes in around 376 IPv4 and 55 IPv6 prefixes, all being originated by different origin ASes and with that single domain, resulting in around 20.5k sibling prefixes mostly with a Jaccard value of 1.
In our dataset, \textit{site24x7.enduserexp.com} is missing on May 2023 and on multiple dates in 2022 and 2021 (see dips of ``diff. org.'' bars), resulting in fewer sibling prefixes for the different organization case.

Moreover, we show the number of \emph{unique} IPv4 and IPv6 prefixes (red and blue lines), which are generally quite stable over time. The slight increase in prefixes over time is due to improvements in OpenINTEL's domain coverage.
Overall, we identify around 6.8k more IPv4 prefixes (46.3k) compared to IPv6 (39.5k) as observed in September 2024.
The high ratio of IPv4 to IPv6 is due to the larger prefix size in IPv6 compared to IPv4, which in turn is able to cover more IP addresses and domains.

To shed more light on sibling prefixes depending on the origin AS organization, \Cref{fig:jacc_over_time_same_diff_orga_28_96} shows the median Jaccard value for sibling prefixes for same and different organizations over time.
The green line shows the median Jaccard value for sibling prefixes of the same organization and has remained relatively stable at 1.0 over the past four years. 
The purple line, the median Jaccard value for sibling prefixes originated by different organizations, also has a Jaccard value of almost one with few dips to around 0.7 and 0.9 in 2021.
The purple line shows an evident influence of the \textit{site24x7.enduserexp.com} domain on median Jaccard values for different organizations.
As such, when that domain is present, it causes many sibling prefixes pairs having a Jaccard value of 1, and its absence is causing the median Jaccard value to drop. %
This indicates that, if we exclude special cases such as Site24x7, sibling prefixes originated by the different organization generally exhibit a high similarity as sibling prefixes originated by the same organization case.
Moreverove, other sites like Catchpoint~\cite{catchpoint}, services that use multi-CDN~\cite{kara2024power, rosenblum2024multi}, alongside, some authoritative DNS that return more than one address to the queries might be potential reasons for different organizations category.

\textbf{To summarize:} We identify around 46k unique IPv4 and almost 39k unique IPv6 prefixes, resulting in more than 76k sibling prefix pairs in the dataset as of September 2024.  
More than half of these sibling prefixes, both the IPv4 and IPv6 prefixes are originated by the same organization, which has a median Jaccard value of 1.
The median Jaccard value for different organizations is sensitive to special case(s) involving a single domain.

\subsection{Business Types}
\label{subsec:sibling_prefixes_business_type}

To better understand the users of sibling prefixes, we analyze the business type of sibling prefix organizations.
For this purpose, we use the most recent ASdb dataset \cite{ziv2021asdb} from January 2024 to identify the business type of each sibling prefix origin AS in our January 2024 sibling prefix snapshot as observed in our dataset.
For this particular analysis, we use only those origin ASes of sibling prefixes that match a single business type, making up around 80\% of all the prefixes.
As we want to focus our attention on business relationships of sibling prefixes originated by different ASes, we therefore exclude  sibling prefix pairs where both IPv4 and IPv6 prefixes have the same origin AS.
The same AS number map to the same business type(s). We also present a more inclusive analysis of the business relationship for sibling prefixes, including those with the same origin AS number in \Cref{subsec:detail_sibling_prefixes_business_type}.

In \Cref{fig:sibling_prefixes_in_ASdb} we show the business type for origin ASes of IPv4 prefixes on the x-axis and for IPv6 prefixes on the y-axis, with the cell color indicating the number of sibling prefixes within this business combination. 
We notice that IT organizations (light yellow cell) form the largest fraction of sibling prefix pairs with more than 10k pairs for IPv4 and IPv6 origin ASes. 
Origin ASes in the education category (black cell in the top left) form the second largest fraction of around 150 sibling prefix pairs.
The vertical and horizontal color density of the cells for IT organizations indicates the usage patterns of sibling prefixes for different ASes, where at least one of the origin ASes belongs to an IT organization. %
Showing that organizations mostly use IT companies, in addition to their own infrastructure for their services (\ie websites).
We also observe a small number of sibling prefix pairs with both origin ASes in the government, media, and retail categories.

\textbf{To summarize:} We observe that 80\% of sibling prefixes with different origin ASes can be mapped to a single business type.
IT companies contribute the highest share of around 10k sibling prefixes, with educational organizations following second.
For most of the remaining sibling prefixes, at least one origin AS belongs to an IT organization.

\begin{figure}[!t]
    \centering
    \includegraphics[width=\linewidth]{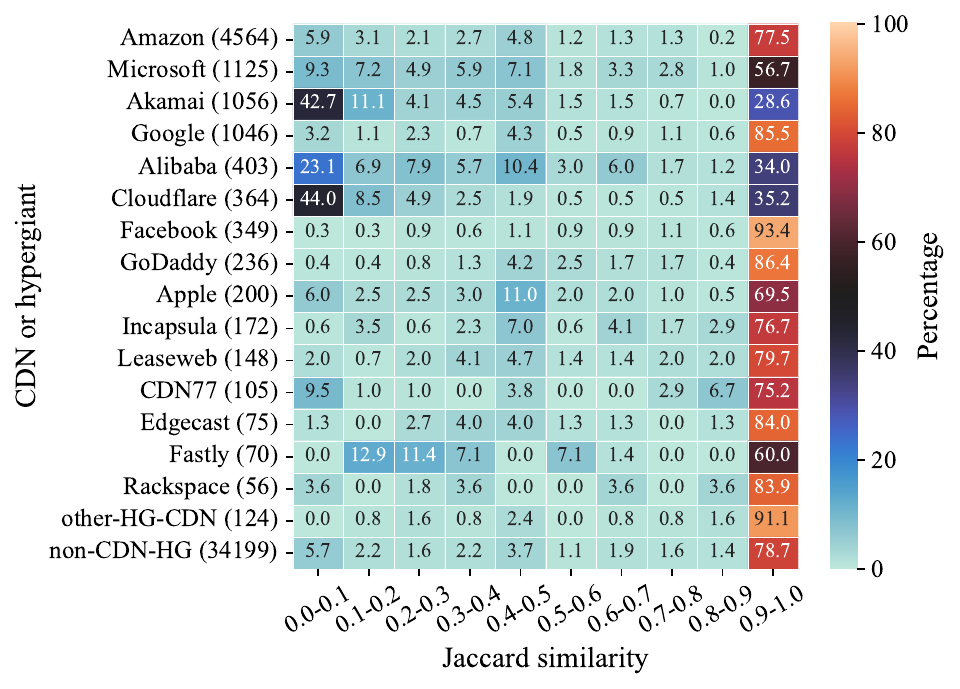}
    \caption{Heatmap of Jaccard similarity values (\%) for sibling prefixes of Hypergiant and CDN networks, and non-CDN-Hypergiant ASes.}
    \label{fig:sp_in_hg_cdn_28_96_v1}
\end{figure}

\subsection{Hypergiants and CDNs}
\label{subsec:sp_in_cdn_and_hgs}

Next, we analyze to what extent sibling prefixes are used by hypergiants (HGs) and content distribution networks (CDNs).
We analyze similarity values for sibling prefixes for known HGs and CDNs and compare them with other networks using our most recent data snapshot of September 11, 2024.
For every prefix pair, we identify their organization using the dataset from Chen \etal \cite{chen2023improving} and use the list of publicly known HGs \cite{bottger2018looking} and CDNs \cite{cdnplanet}.
Note, we only consider SP-Tuner output sibling prefixes where IPv4 and IPv6 prefixes belong to the same organization of HG or CDN.

\Cref{fig:sp_in_hg_cdn_28_96_v1} shows the distribution of Jaccard similarity values in HGs, CDNs, and other networks. 
The y-axis shows the organization names of one of the 24 HGs or CDNs, with the number in parentheses indicating the count of sibling prefixes for this specific HG or CDN.
We group all the HGs and CDNs that contribute less than 50 sibling prefix pairs as ``other-HG-CDN''. 
Moreover, sibling prefixes that are neither originated by HGs nor CDNs are shown as ``non-CDN-HG'' in the bottom row.

The x-axis shows the Jaccard similarity values in ten cells, ranging from the smallest Jaccard similarity value of 0.0--0.1 to a very high similarity of 0.9--1.0.
The color and number in each heatmap cell indicate the percentage of sibling prefix pairs for each HG or CDN with specific Jaccard similarity value.
Thus, each row shows the distribution of sibling prefixes for a specific HG or CDN, summing up to 100\%.

Examining the number of sibling prefixes, we identify the the highest number of sibling prefixes for Amazon, with 4564 pairs, followed by Microsoft, Akamai, and Google with 1125, 1056, and 1046 sibling prefix pairs, respectively.
The right-most (0.9--1.0) column, with a few exceptions, contains the highest shares of sibling prefixes.
This shows that many sibling prefixes in an HG and CDN are very similar in containing perfect matching domains.
Cloudflare and Akamai have the highest share of sibling prefixes in the worst matching category, \ie with Jaccard values of 0.0--0.1.
This aligns with certain CDNs moving semantics from IP addresses to domain names \cite{majkowski2022cloudflare, marwan2021ties, Larisch2024topaz}.
For non-CDN-HG ASes, 78\% of sibling prefixes have a high Jaccard similarity value of 0.9--1.0.
We observe the highest matching category for 93\% of Facebook sibling prefixes and 91\% of other-HG-CDN category followed by GoDaddy having 86\% as the top three.
Overall, we see that CDNs and HGs exhibit high similarity for sibling prefixes.%

\textbf{To summarize:} We identify 24 hypergiant or CDNs using sibling prefixes. 
Amazon, Microsoft, and Akamai are the top three HGs with the highest number of sibling prefixes.
The vast majority of sibling prefixes in HGs and CDNs are either very similar, with few HGs being exceptions with low similarity scores.

\begin{figure}[!t]
    \centering
    \includegraphics[width=\linewidth]{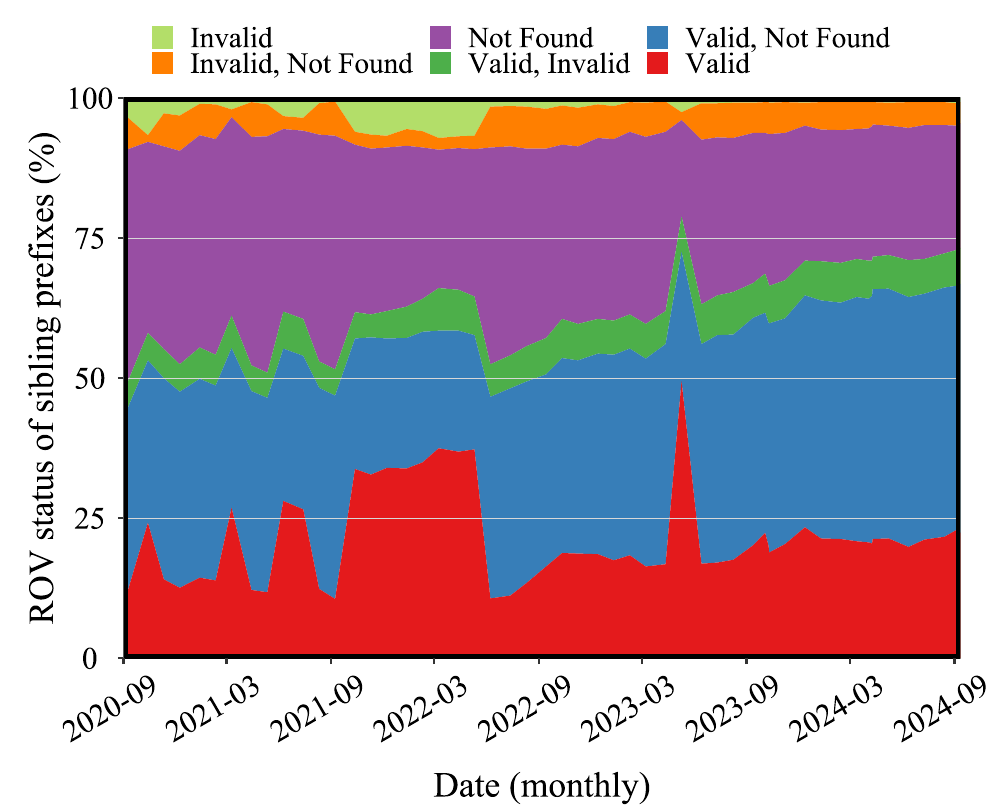}
    \caption{Route Origin Validation (ROV) status of sibling prefix pairs in the RPKI data over time.}%
    \label{fig:sps_in_rpki}
\end{figure}

\subsection{RPKI Validity}
\label{subsec:DS_prefix_rov}

The resource public key infrastructure (RPKI) can be used to check the legitimacy of a prefix announcement in BGP.
Thus, we use RPKI data \cite{rpkidata} to analyze the legitimacy of sibling prefix pairs as observed in our dataset.
For this analysis we use the BGP-announced prefix sizes instead of the SP-Tuner output, as those align better for this BGP-specific analysis.

\Cref{fig:sps_in_rpki} illustrates the route origin validation (ROV) status for sibling prefixes from monthly snapshots between September 2020 and September 2024.
While the x and y axes show the date and percentage, the share for each ROV status is displayed as the area over time.
The red color indicates that both prefixes in a sibling prefix pair have a valid ROV status.
The blue indicates that at least one of two prefixes has a valid ROV status, and the other was not found in RPKI.
We find that for nearly 50\% of sibling prefixes in 2020, to around 65\% of sibling prefixes in September 2024, either one (\ie the IPv4 or IPv6 prefix) or both prefixes of a sibling pair have a valid ROV status.
For a small fraction of sibling prefixes (2--8\%), we observe a conflicting ROV status, \ie one prefix has a valid and the other an invalid ROV status (green).
The share of both prefixes not being in the RPKI (purple) decreases from 40\% in September 2020 to almost 20\% in September 2024.
For around 10\% of sibling prefixes, we observe an ROV status of either one prefix being invalid and another not found or both prefixes having an invalid ROV status, as shown in orange and light green colors, respectively.

These findings are in line with the general trend of increased RPKI deployment in the Internet \cite{rpkimonitor} and can help protect sibling prefixes from routing incidents such as BGP hijacks \cite{bgpmon, caida_hijacks_observatory, qin2022themis}.
For all sibling prefix pairs, where one prefix is valid, and the other is not found, it is crucial to add the second prefix to the RPKI by creating a valid route origin authorization (ROA) object.
It is even more important for cases where entire sibling prefix pair have invalid, a combination of (valid, invalid) or (invalid, not found)  ROV state.
In such cases, traffic might not be routed to those prefixes over both IP protocol versions, leading to a lack of resilience.

\textbf{To summarize:} For around 65\% of sibling prefixes, at least one sibling prefix has a valid ROV status. Still, for nearly 10\% of sibling prefixes, at least one prefix has an invalid ROV status.

The decrease in number of not-found ROV cases over time, is in line with the overall increase in RPKI deployment.
Creating valid ROV status for both pairs of sibling prefixes is essential to protect sibling prefixes and services running on those prefixes from routing incidents like BGP hijacks.
\section{Related Work}
\label{sec:reated_work}

Previous studies explored various aspects of siblings, alias resolution, fingerprinting techniques, IP addresses, and device detection on the Internet, focusing on the IP address level rather than on the prefix level.

\parx{Server siblings:}
In 2013, Berger \etal \cite{berger2013internet} studied the associations between Internet DNS client resolver IPv4 and IPv6 addresses. The study uses passive and active techniques to identify 674k associated address pairs. 
In 2015, Beverly and Berger \cite{beverly2015server} identified pairs of IPv4 and IPv6 server addresses potentially assigned to the same physical machine as siblings using TCP-reachable devices. 
In 2016, Czyz \etal \cite{Czyz2016DontFT} explored potential policy discrepancies between IPv4 and IPv6 dual-stack hosts and found that a ports are nearly always more open in IPv6 compared to IPv4.
In 2017, Scheitle \etal \cite{scheitle2017large} identified pairs of server IPv4 and IPv6 addresses as siblings using TCP timestamps. 
They used manually crafted algorithms and machine-learned decision trees to classify pairs of IPv4 and IPv6 server addresses as siblings, \ie running on the same machine. 
In 2021 and 2023, Albakour \etal \cite{albakour2021third,albakour2023pushing} explored SNMPv3, SSH, and BGP for remotely fingerprinting network infrastructure in the wild. 
By sending unsolicited and unauthenticated SNMPv3 requests, the authors identified the status of network devices, including device vendor and uptime. 

\parx{Alias resolution and fingerprinting:} 
In 2012, Keys \etal \cite{keys2012internet} introduced MIDAR, a tool leveraging the IP ID for alias resolution.
In 2013, Luckie \etal \cite{luckie2013speedtrap} proposed Speedtrap, a technique that induces fragmented IPv6 responses from router interfaces in a particular temporal pattern to for router fingerprinting.
In 2015, Padnamabhan \etal \cite{padmanabhan2015uav6} introduced UAv6, a new alias resolution technique for IPv6. They probed the unused addresses using ICMPv6 Address Unreachable responses for IPv6 alias resolution. 
In 2016, Beverly and Berger \cite{beverly2013ipv6}, focused on discovering router-level IPv6 topologies. They used fingerprinting-based IPv6 alias resolution techniques to induce fragmented responses from IPv6 router interfaces. 
In 2020, Vermeulen \etal \cite{vermeulen2020alias} presented a tool called ``Limited Ltd.'' which exploits ICMP rate limiting for alias resolution.

\parx{Address assignment:}
In 2020, Padnamabhan \etal \cite{padmanabhan2020dynamips} looked into the IPv6 address assignments in various networks and how they relate to IPv4 dynamics. 
They found IPv6 assignments having longer durations than IPv4, often remaining stable for months.
In 2023, Rye and Levin \cite{rye2023ipv6} leveraged NTP pools to gather 7.9 billion IPv6 addresses from NTP clients, examining the potential benefits and harm of larger IPv6 hitlists and the possibility of measurement and analysis.
In the same year, Hsu \etal \cite{hsu2023fiat} identified the most common delegation prefix length by RIRs being /32, and the most prevalent BGP prefix length in IPv6 being a /48.

\parx{Prefix lists:}
The work most closely related to ours is by Naab \etal \cite{naab2019gaining}, who introduced the concept of prefix top lists in 2019. After aggregating domain-based top lists into network prefixes, they leveraged a Zipf distribution to assign weights to each prefix. 
They found that different domain-based top lists provide differentiated views on the Internet prefixes with minimal weight change over time.

To the best of our knowledge, our work is the first to investigate sibling relationships between IPv4 and IPv6 on a prefix-level.
\section{Discussion}
\label{sec:discussion}

\noindent\textbf{Sibling prefix input dataset:}
In this research, we introduce a technique to identify sibling prefixes in the Internet.
While our analysis uses domain similarity as input to identify sibling prefixes, we argue that we can identify sibling prefixes using other services, such as DNS MX records, rDNS names, or aliased hosts.
We cross-check our findings with real world deployments of RIPE Atlas probes as a ``ground-truth'' dataset, finding a high overlap.
Additionally, we perform port scans and find correlations between sibling prefixes identified using DNS and port scan datasets.
An even larger and more comprehensive input dataset might allow to further increase the number of identified sibling prefixes in the Internet.

\noindent\textbf{Choosing the right prefix size:}
With SP-Tuner we present an algorithm to fine-tune the CIDR size of sibling prefixes to achieve even higher Jaccard similarity values.
Although we mostly use the /28 and /96 pairs as the SP-Tuner thresholds in this study, options in selecting any pair of thresholds make SP-Tuner flexible, and allow to choose appropriate prefix size thresholds based on the need to further fine-tune the resulting sibling prefixes.
While this is a good first step in finding well suited prefix sizes, it is not a silver bullet.
IPv4 address exhaustion \cite{prehn2020wells,qin2022themis} leads to smaller and smaller prefixes being allocated by RIRs, which in turn then leads to more and more fragmentation of the IPv4 address space.
Furthermore, identifying the ``right'' prefix size remains a challenge, especially in IPv6 \cite{prefixlendraft}.
These factors make it exceedingly challenging to identify sibling prefixes.
Therefore, it might be useful to look into sibling prefix \emph{set} pairs, \ie a set of IPv4 prefixes which are siblings of a set of IPv6 prefixes.
This could alleviate challenges such as address space fragmentation by pairing different IPv4 fragments with their IPv6 counterpart.
We leave the analysis of sibling prefix set pairs to future work.

\noindent\textbf{Domains instead of addresses:}
Moreover, some network architectures have shifted away from owning dedicated IP addresses for their services, instead sharing the same IP prefixes for their services and across multiple data centers \cite{majkowski2022cloudflare,hustonipv6}.
Therefore, identifying sibling prefixes in the Internet will help network operators around the globe to make more informed decisions, such as applying similar routing policies for a set of domains running on IPv4 and IPv6 as sibling prefix pairs.
Sibling prefixes enable to allow, block, filter, or otherwise differentiate a group of domains collectively based on their prefixes.
One example application of sibling prefixes is the adaption of IPv4 spam blocklists to IPv6, which closes the backdoor for spammers to switch to IPv6 if they are blocked on IPv4.
Furthermore, a list of sibling prefixes with high Jaccard values could help the research community to more easily adapt their IPv4 analysis technique to IPv6.
Therefore, we plan to regularly publish a list of sibling prefixes to be used by network operators and researchers at \href{https://sibling-prefixes.github.io}{sibling-prefixes.github.io}.
\section{Conclusion}
\label{sec:conclusion}

In this paper, we provided a thorough analysis of sibling prefixes in the Internet by applying the Jaccard similarity index on dual-stack domains.
We introduced the SP-Tuner algorithm that identifies and fine-tunes IP prefix sizes resulting in tailored sibling prefixes having higher Jaccard values and smaller CIDR sizes.
With SP-Tuner we were able to improve the share of perfect match sibling prefixes from 52\% to 82\%.
We found that more than half of sibling prefixes belong to the same organizations, with the majority being associated with IT companies.
We found sibling prefixes to be relatively stable over time and to be prevalent in 24 hypergiant and CDN networks.
Sibling prefixes showed similar RPKI adoption rates as other prefixes, we did however find instances with inconsistent ROV status between IPv4 and IPv6 prefixes of sibling pairs.
Finally, we plan to regularly publish sibling prefix lists for researchers and operators.
\section*{Acknowledgements}
We thank the anonymous reviewers for their valuable feedback, as well as our shepherd, Todd Arnold.
We also thank the OpenINTEL team for providing the DNS data and the RIPE Atlas team for providing the probe data.
Part of this work was carried out during Fariba's internship at the National Institute of Informatics, Japan (NII). %
\bibliographystyle{ACM-Reference-Format}
\balance
\bibliography{paper}
\newpage
\appendix
\clearpage

\begin{figure}[t]
  \begin{minipage}{0.45\textwidth}
      \begin{algorithm}[H]
        \caption{SP-Tuner-LS (Less Specific)}\label{alg:sp_tuner_ls_algorithm}
        \scriptsize
        \begin{algorithmic}[2]
          \Statex \textbf{Data:} DS domains, IPv4\_addresses, IPv6\_addresses, sibling prefix pairs, Jaccard similarity
            \Statex \textbf{Result:} Refined sibling prefix pairs with improved Jaccard similarity
            \Statex \textbf{Initialization:}
            \Statex \hspace{\algorithmicindent} $\textit{tree\_v4} \gets \{\textit{DS\_domains}, \textit{IPv4\_addresses}\}$
            \Statex \hspace{\algorithmicindent} $\textit{tree\_v6} \gets \{\textit{DS\_domains}, \textit{IPv6\_addresses}\}$
            \Statex \hspace{\algorithmicindent} $\textit{curr\_jacc} \gets \textit{precomputed non-zero value}$ %
            \Statex \hspace{\algorithmicindent} $\textit{prefix\_v4\_len\_thresh}$ \Comment{IPv4 prefix length threshold: 1 levels up}
            \Statex \hspace{\algorithmicindent} $\textit{prefix\_v6\_len\_thresh}$ \Comment{IPv6 prefix length threshold: 4 levels up}
            \For{$(\textit{sibling\_prefix\_v4,\ sibling\_prefix\_v6}) \in \textit{sibling\_prefix\_pairs}$} 
            \State $\textit{tree\_v4} \gets \textit{sibling\_prefix\_v4}$
            \State $\textit{tree\_v6} \gets \textit{sibling\_prefix\_v6}$
            \State $\textit{new\_jacc} \gets 0$
            \While{$\textit{new\_Jacc} \leq \textit{curr\_Jacc}$}
              \State $\textit{prefix\_v4\_superprefixes} \gets \textit{GetNextSuperprefixes}(\textit{sibling\_prefix\_v4})$ %
              \State $\textit{prefix\_v6\_superprefixes} \gets \textit{GetNextSuperprefixes}(\textit{sibling\_prefix\_v4})$ %
              \For{$(\textit{prefix\_v4}, \textit{prefix\_v6}) \in \{ \textit{prefix\_v4\_superprefixes}, \textit{prefix\_v6\_superprefixes} \}$}
                \State $\textit{new\_Jacc} \gets \text{max}(Jaccard(\textit{prefix\_v4}, \textit{prefix\_v6}))$
                \If{$(\textit{IsASnumChange(prefix\_v4)})$ || $(\textit{IsASnumChange(prefix\_v6)})$}
                  \State \textbf{\textit{return}} $(\textit{prefix\_v4}, \textit{prefix\_v6})$
                  \ElsIf{$(\textit{prefix\_v4\_len} \leq \textit{v4\_len\_thresh})$ \\
                    \hspace{2em} || $(\textit{prefix\_v6\_len} \leq \textit{v6\_len\_thresh})$}
                      \State \Return $(\textit{prefix\_v4}, \textit{prefix\_v6})$

                \EndIf
              \EndFor
            \EndWhile
            \EndFor
        \end{algorithmic}
      \end{algorithm}
  \end{minipage}
\end{figure}

\begin{figure*}[!t]
  \minipage[t]{0.48\textwidth}
    \centering
    \includegraphics[width=1.05\linewidth]{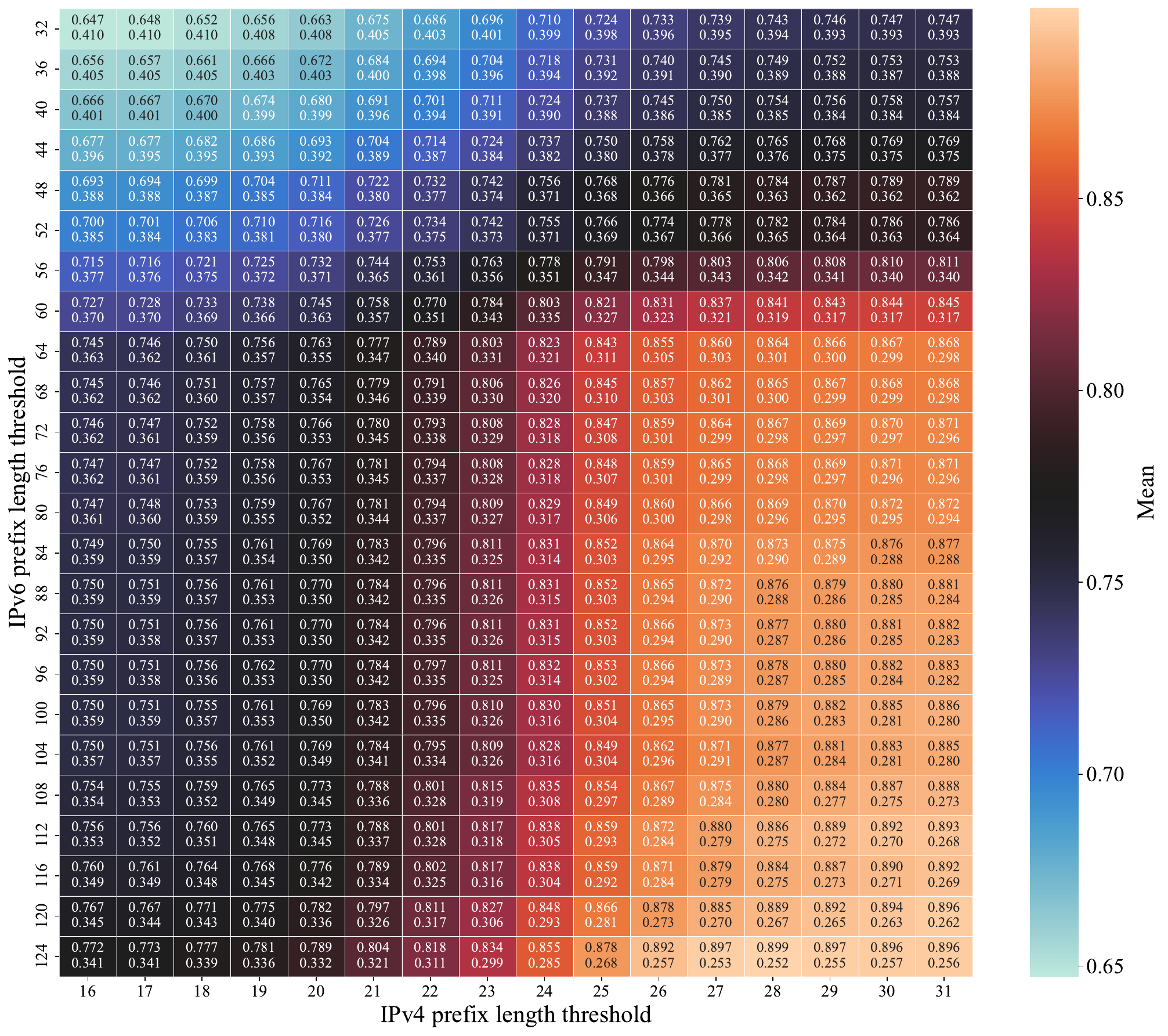}
    \caption{Heatmap of SP-Tuner Algorithm Performance: Mean Jaccard Index (top) and Standard Deviation (bottom) across IPv4 (x-axis) and IPv6 (y-axis) CIDR size thresholds.}
      \label{fig:sp_tuner_mean_std_all}
    \endminipage
\end{figure*}

\section{Appendix}
\label{sec:appendix}
\subsection{SP-Tuner-LS (Less Specific)}
\label{subsec:sp_tuner_ls}

Algorithm \ref{alg:sp_tuner_ls_algorithm} with the name SP-Tuner-LS shows the traversal of each tree in the upward direction from the point of prefix inserts. 
At every step, it generates the supernet of the current prefix and checks Jaccard similarity until improvement occurs. 
Another point to check at every step of going up is the AS number; this might happen when generating the supernet of a prefix, causing the originating AS number to change. 
To address this, leverage the RouteView data for that specific date, ensuring the same date as our input data.

\begin{figure*}[!t]
  \minipage[t]{0.38\textwidth}
  \centering
  \includegraphics[width=\linewidth]{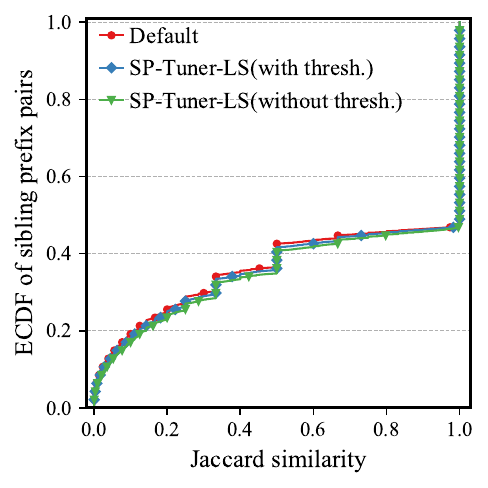}
  \caption{Jaccard similarity after applying SP-Tuner-LS on sibling prefix pairs data.}
  \label{fig:sp_tuner_ls_2}
\endminipage
\hfill
  \minipage[t]{0.50\textwidth}
    \centering
    \includegraphics[width=\linewidth]{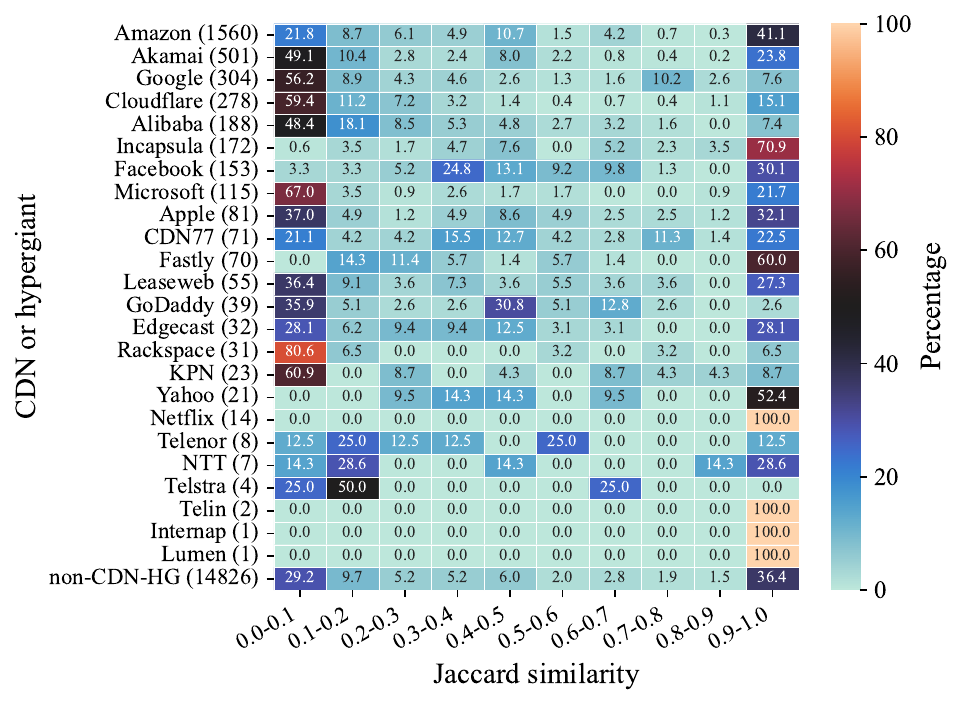}
    \caption{Heatmap of Jaccard similarity for the CDN, HGs and non-CDN-HG for the default case.}
    \label{fig:sp_in_hg_cdn-2}
  \endminipage
\end{figure*}

The \Cref{fig:sp_tuner_ls_2} shows the results of applying SP-Tuner-LS. 
Using this approach and increasing the prefix sizes for fine-tuning the sibling prefixes does not improve the Jaccard similarity values, as depicted by the green line in \Cref{fig:sp_tuner_ls_2}. 
However, the blue line indicates that setting a threshold on the prefix sizes for how much they could be increased does not result in a significant improvement in Jaccard similarity.

\subsection{SP-Tuner-MS Sensitivity Analysis}
\label{subsec:sp_tuner_sensitivity}
To achieve an improved Jaccard similarity, we conduct a sensitivity analysis to evaluate the behavior of SP-Tuner-MS under various prefix length thresholds.
The results of these tests, performed on IPv4 prefixes ranging from /16 to /31 and IPv6 prefixes ranging from /32 to /124, is illustrated in \Cref{fig:sp_tuner_mean_std_all}.
In each cell, the top value represents the mean Jaccard similarity across all sibling prefixes, while the bottom value indicates the standard deviation of the Jaccard values.

\subsection{Detail Analysis of Hypergiants and CDNs of Sibling Prefixes}
\label{subsec:detail_sibling_prefixes_HG_CDN}
\begin{figure*}[!t] 
     \minipage[t]{0.48\textwidth}
    \centering
    \includegraphics[width=\linewidth]{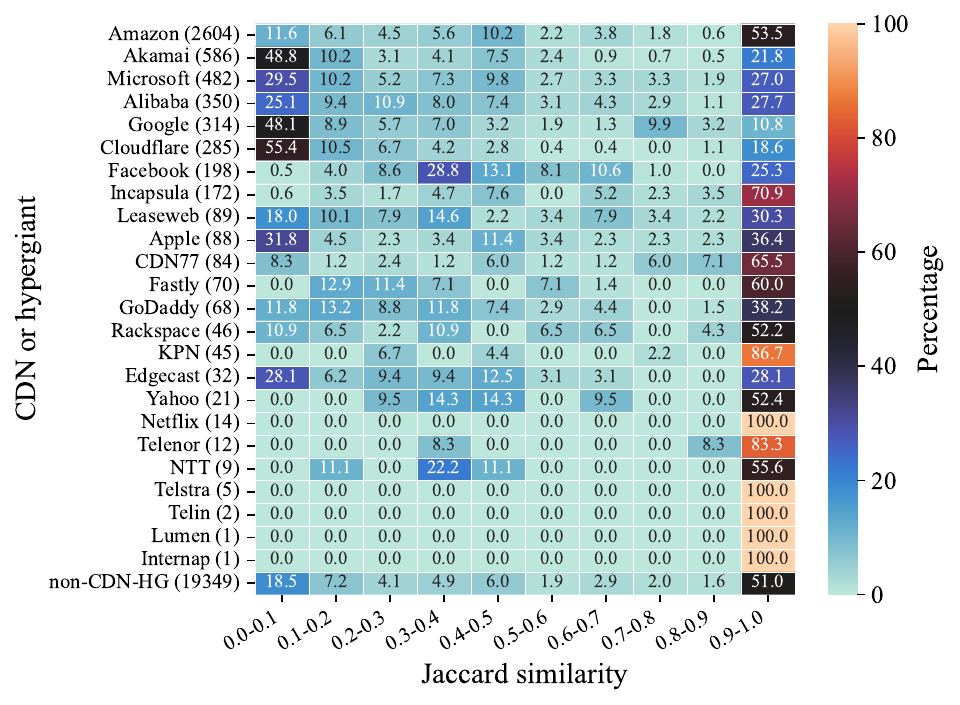}
    \caption{Heatmap of Jaccard similarity for the CDN, HGs and non-CDN-HG for the SP-Tuner thresholds of /24-/48.}
    \label{fig:sp_in_hg_cdn_24_48-2}
  \endminipage
  \hfill
  \minipage[t]{0.48\textwidth}
    \centering
    \includegraphics[width=\linewidth]{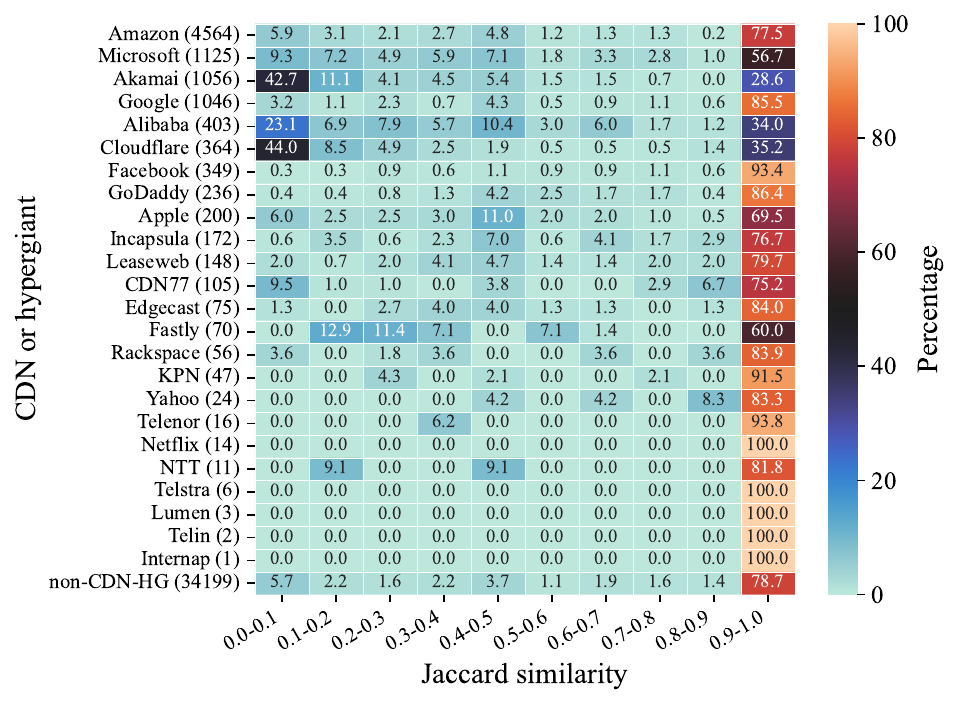}
    \caption{Heatmap of Jaccard similarity for the CDN, HGs and non-CDN-HG for the SP-Tuner thresholds of /28-/96.}
    \label{fig:sp_in_hg_cdn_28_96_2}
  \endminipage
  \hfill
\end{figure*}

\Cref{fig:sp_in_hg_cdn-2,fig:sp_in_hg_cdn_24_48-2} and \Cref{fig:sp_in_hg_cdn_28_96_2} shows a detaild information of distribution of Jaccard similarity values for sibling prefixes in HGs, CDNs, for the default case as observed in BGP, using /24-/48 thresholds, and the /28-/96 thresholds of SP-Tuner algorithm.

\subsection{Detail Analysis of Business Type of Sibling Prefixes}
\label{subsec:detail_sibling_prefixes_business_type}

As a detailed information on the business types of the origin ASes of sibling prefixes explained in \Cref{subsec:sibling_prefixes_business_type}, we consider the following two cases:

\subsubsection{Origin AS of Sibling Prefixes for Business Type Analysis } 
\label{subsubsec:sibling_prefixes_origins_only_business_type}

In this case, instead of counting the number of sibling prefixes, we count the number of unique origin AS pairs of the sibling prefixes. 
This analysis helps overcome the possible influence of a few origin ASes that might contribute to many sibling prefixes. 
While we still have both the ASdb and sibling prefixes data from January 2024, we exclude cases where the IPv4 and IPv6 prefixes of a sibling prefix pair have the exact origin AS number. 
\Cref{fig:sibling_prefixes_origins_in_ASdb} illustrates the number of origin AS pairs for various business types of the origin ASes.
The x-axis represents the business type of the IPv4 prefix origin AS, and the y-axis represents the business type of the IPv6 prefix origin AS.

The overall heatmap pattern for the count of origin AS pairs of sibling prefixes is very similar to the \Cref{fig:sibling_prefixes_in_ASdb} in \Cref{subsec:sibling_prefixes_business_type} explained for the number of sibling prefixes. 
As it can be seen in \Cref{fig:sibling_prefixes_origins_in_ASdb}, the largest pair of ASes, more than one thousand pairs shown in light yellow color, are the cases where both origin ASes are It organizations. 
The origin AS pairs having an Education business type are the second largest group or origin AS pair for sibling prefixes. 
The row and column of cells for the IT business are colored, while the majority of other cells are empty, indicating that at least one of the origin ASes for the sibling prefixes is an IT organization. 
We can conclude that even if we consider a pair of unique origin ASes for the sibling prefixes in the business type analysis, the overall takeaway of the business type explained in \Cref{subsec:sibling_prefixes_business_type} still holds.

\subsubsection{Unfiltered Sibling Prefixes for Business Type Analysis} 
\label{subsubsec:unfiltered_sibling_prefixes_business_type}
Finally, we do not filter the cases where both prefixes in a sibling prefix pair have the same origin AS number. 
The idea is to provide a more inclusive picture of the sibling prefixes even if both origin AS have the same AS number.
We want to understand the business types of those origin ASes and how much this affects the analysis we explained \Cref{subsec:sibling_prefixes_business_type}.
 
\begin{figure*}[!t]
  \minipage[t]{0.46\textwidth}
\centering
\includegraphics[width=1.05\linewidth]{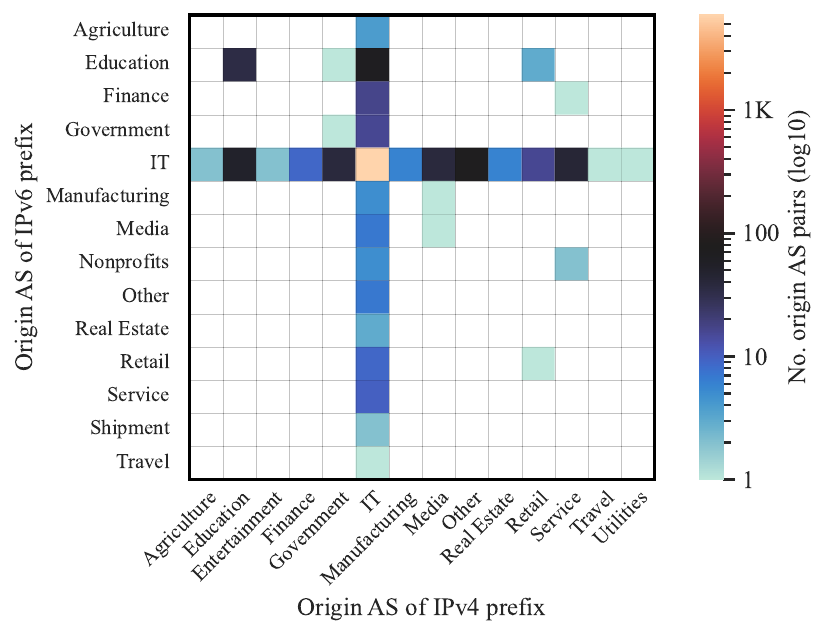}
\caption{Business types heatmap for origin ASes of sibling prefixes, with IPv4 on x-axis and IPv6 on y-axis, indicating the count of origin AS pairs in each cell. }
  \label{fig:sibling_prefixes_origins_in_ASdb}
\endminipage
  \hfill
  \minipage[t]{0.48\textwidth}
\centering
\includegraphics[width=\linewidth]{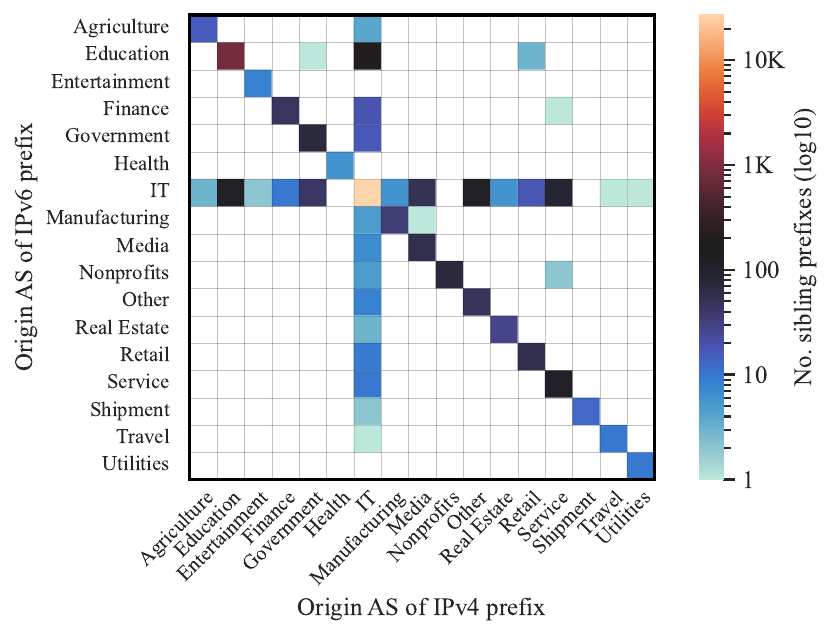}
\caption{Business types heatmap for origin ASes of sibling prefixes, with IPv4 on x-axis and IPv6 on y-axis, indicating the count of sibling prefix pairs in each cell. }
  \label{fig:all_sibling_prefixes_origins_in_ASdb}
\endminipage
\end{figure*}

\Cref{fig:all_sibling_prefixes_origins_in_ASdb} shows the business type analysis for all the sibling prefixes as observed in January 2024. 
We notice that the highest number in the color bar on the right side of the plot is increased from the range of one thousand, shown as 1K, to more than ten thousand, shown as 10K. 
The high number of sibling prefix pairs is because many sibling prefixes with the exact origin AS number are included in the data and as explained in \Cref{subsec:evolution_of_sibling_prefixes}.
Another significant difference in \Cref{fig:all_sibling_prefixes_origins_in_ASdb} comparing to \Cref{fig:sibling_prefixes_origins_in_ASdb} and \Cref{fig:sibling_prefixes_in_ASdb} in \Cref{subsec:sibling_prefixes_business_type} is the diagonal line of colorful cells. 
The sibling prefix pairs having the exact origin AS fall into the diagonal line, as both origin AS numbers fall into the same type of business. As a result, we see the diagonal line of the colorful cell in the heatmap in \Cref{fig:all_sibling_prefixes_origins_in_ASdb}.
However, the cell having the highest number of sibling prefixes, more than ten thousand sibling prefix pairs, shown in light yellow color, still belongs to the IT organizations, which are high in number but similar to the previous analysis of the business type of the origin ASes for sibling prefixes.
The IT organization has an overall higher number of sibling prefixes on the corresponding row and column, indicating that one of two origin ASes of sibling prefixes falls into the IT business type. The result is similar to all previous analyses concerning the business type of the origin ASes for sibling prefixes.

\subsection{Over Time Analysis of Sibling Prefixes}
\label{subsec:over_time_analysis_of_siblings}

\begin{figure*}[!t]
  \minipage[t]{0.31\textwidth}
    \centering
    \includegraphics[width=\linewidth]{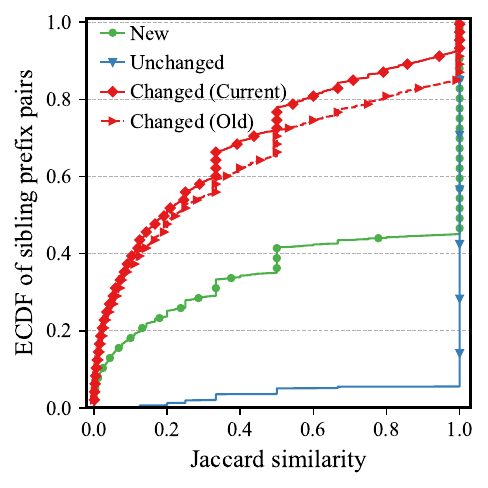}
    \caption{Changes of Jaccard similarity index value over time for the default case.}
    \label{fig:ecdf_jacc_overtime_default}
  \endminipage
  \hfill
  \minipage[t]{0.31\textwidth}
    \centering
    \includegraphics[width=\linewidth]{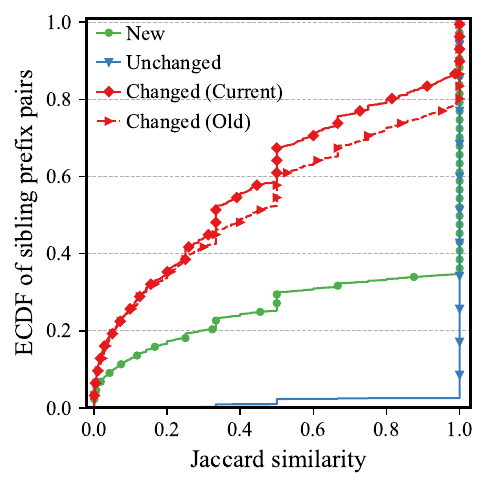}
    \caption{Changes of Jaccard similarity index value over time for SP-Tuner with /24-/48 thresholds.}
    \label{fig:ecdf_jacc_overtime_24_48}
  \endminipage
  \hfill
  \minipage[t]{0.31\textwidth}
    \centering
    \includegraphics[width=\linewidth]{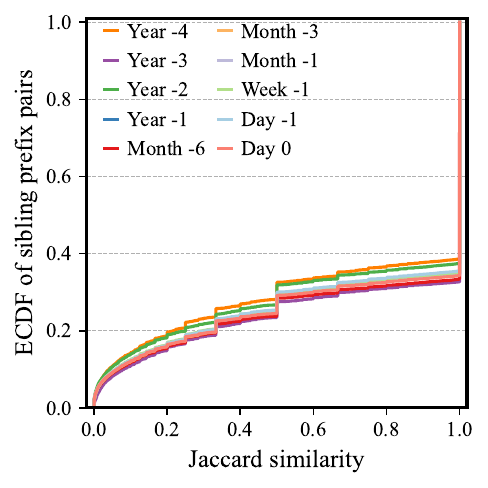}
    \caption{Impact of SP-Tuner with 24-48 thresholds on Jaccard similarity values on various snapshots over time.}
    \label{fig:eJaccard_over_time_analysis_24_48}
  \endminipage
\end{figure*}

\Cref{fig:ecdf_jacc_overtime_default} illustrates over time Jaccard similarity changes for sibling prefixes for the default case, \Cref{fig:ecdf_jacc_overtime_24_48} for the SP-Tuner with a /24-/48 threshold case. On the other hand, \Cref{fig:eJaccard_over_time_analysis_24_48} illustrates the impact of the SP-Tuner algorithm with a /24-/48 threshold on Jaccard similarity of sibling prefixes for various data snapshots over time. 

\subsection{Origin ASes of Sibling Prefixes}
\label{subsec:origin_ASes_of_siblings}

\begin{figure*}[!t]
  \minipage[t]{0.48\textwidth}
    \centering
    \includegraphics[width=\linewidth]{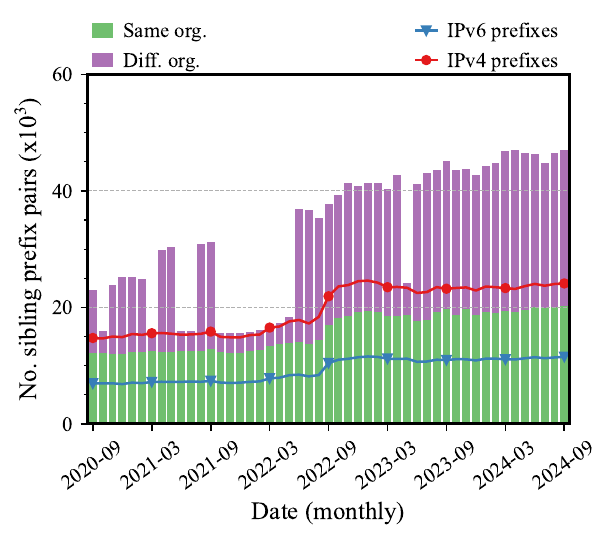}
    \caption{Sibling prefixes (bars) over time in same and different organizations for default case with lines showing the number of unique IPv4 and IPv6 prefixes over time.}
    \label{fig:bar_asnum_org_default}
  \endminipage
  \hfill
  \minipage[t]{0.48\textwidth}
    \centering
    \includegraphics[width=\linewidth]{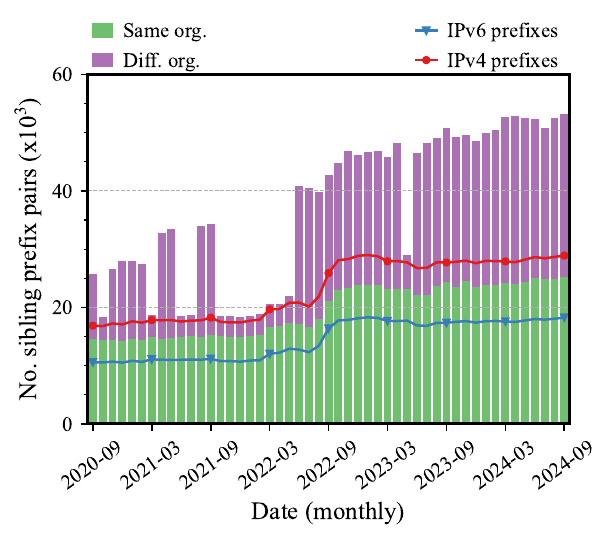}
    \caption{Sibling prefixes (bars) over time in same and different organizations for SP-Tuner with /24-/48 thresholds, lines showing the number of unique IPv4 and IPv6 prefixes over time.}
    \label{fig:bar_asnum_org_24_48}
  \endminipage
\end{figure*}

\begin{figure*}[!t]
    \minipage[t]{0.48\textwidth}
    \centering
    \includegraphics[width=\linewidth]{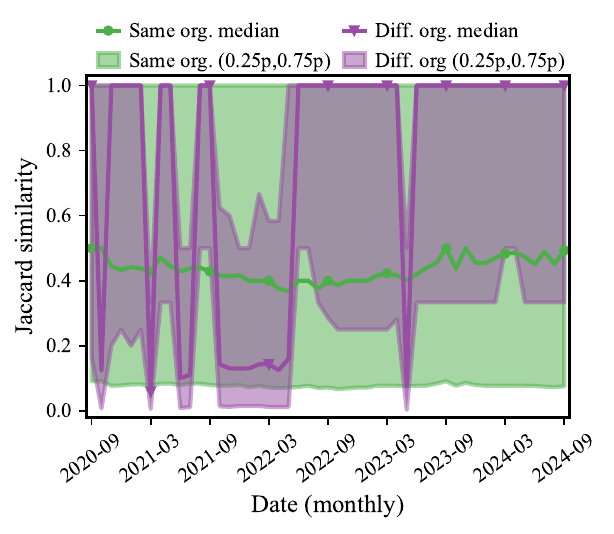}
    \caption{Median Jaccard for same and different organization over time for default case.}
    \label{fig:jacc_over_time_same_diff_orga_default}
  \endminipage
  \hfill
  \minipage[t]{0.48\textwidth}
  \centering
  \includegraphics[width=\linewidth]{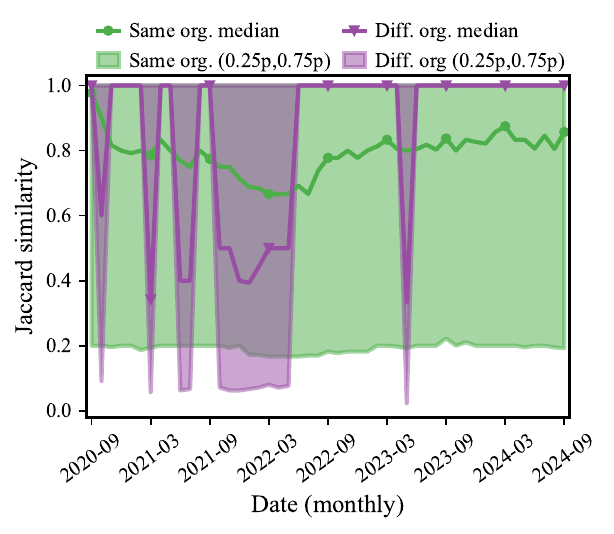}
  \caption{Median Jaccard for same and different organization over time for SP-Tuner with /24-/48 thresholds.}
  \label{fig:jacc_over_time_same_diff_orga_24_48}
\endminipage
\end{figure*}

\Cref{fig:bar_asnum_org_default,fig:bar_asnum_org_24_48} depict the number of sibling prefixes over time for the same and different organizations in bars, the lines present the number of unique IPv4 and IPv6 prefixes for the default case and the SP-Tuner case with /24-/48 thresholds. On a similar order of sibling prefixes for the default and SP-Tuner case, \Cref{fig:jacc_over_time_same_diff_orga_default,fig:jacc_over_time_same_diff_orga_24_48} show the median Jaccard value changes over time for both cases.  


\subsection{Domains and CIDR sizes of Sibling Prefixes}
\label{subsec:domains_and_cidrs_of_sibling_prefixes}
\begin{figure*}[!t]
  \minipage[t]{0.48\textwidth}
    \centering
    \includegraphics[width=\linewidth]{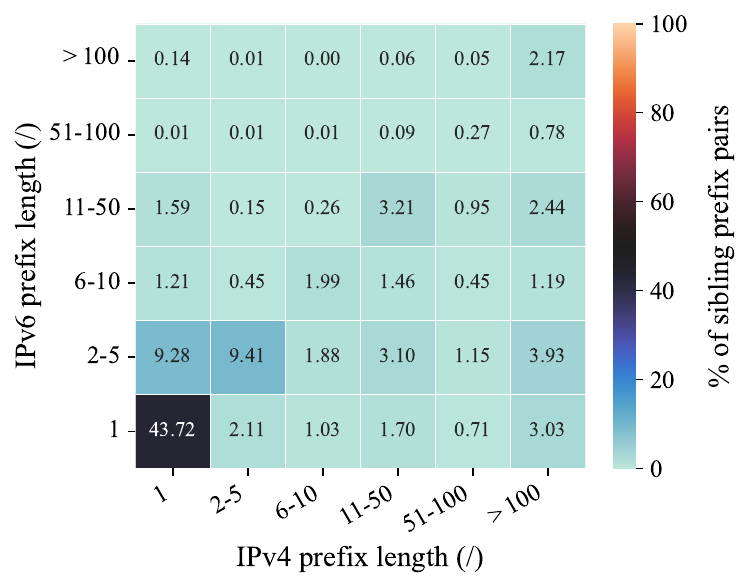}
    \caption{Heatmap of sibling prefixes \% based on number of domains for default case.}
    \label{fig:sp_domain_category_p}
  \endminipage
  \hfill
  \minipage[t]{0.48\textwidth}
    \centering
    \includegraphics[width=\linewidth]{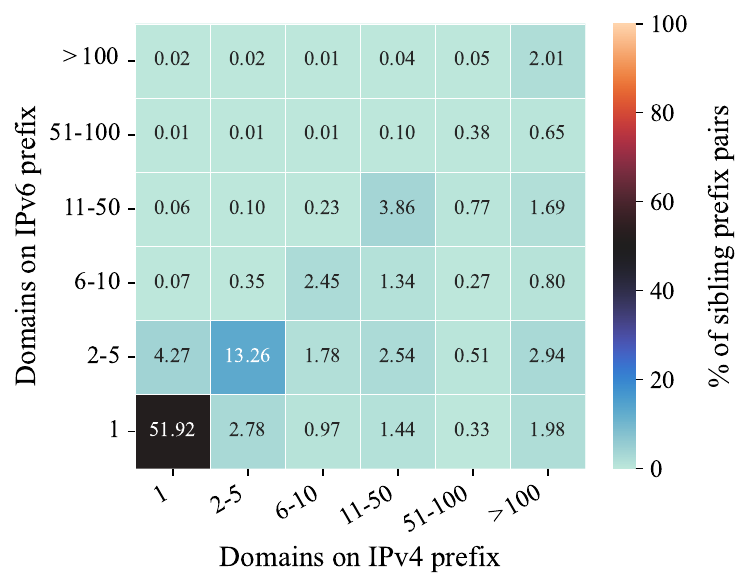}
    \caption{Heatmap of sibling prefixes \% based on number of domains for SP-Tuner with 24-48 thresholds.}
    \label{fig:sp_domain_category_24_48_p}
  \endminipage
\end{figure*}

\begin{figure*}[!t]
  \minipage[t]{0.48\textwidth}
    \centering
    \includegraphics[width=\linewidth]{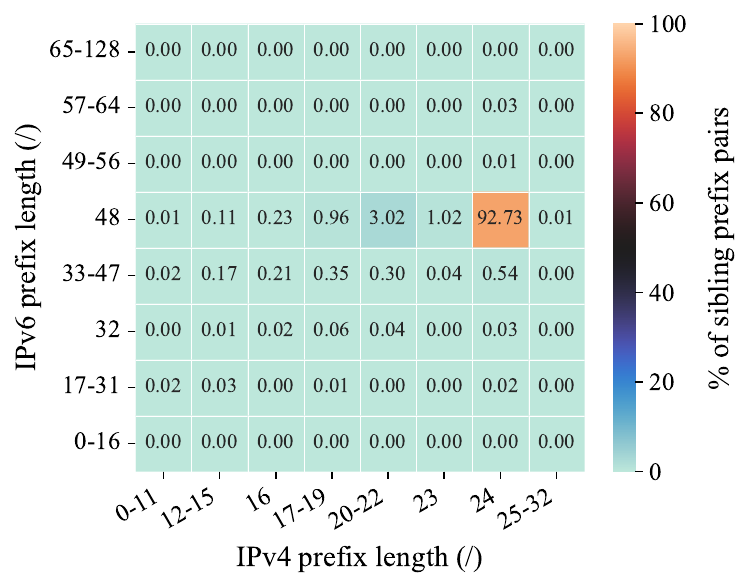}
    \caption{Heatmap of sibling prefixes CIDR size \% for SP-Tuner with /24-/48 thresholds.}
    \label{fig:sp_prefix_size_24_48_p}
  \endminipage
  \hfill
  \minipage[t]{0.48\textwidth}
    \centering
    \includegraphics[width=\linewidth]{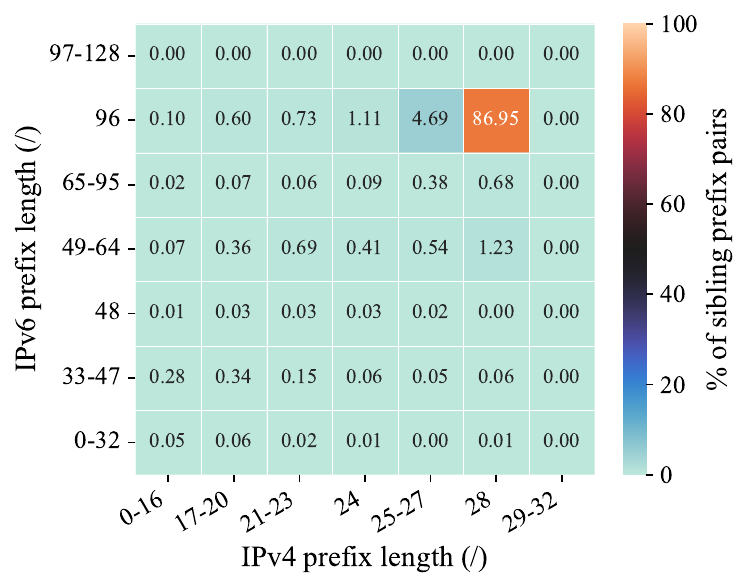}
    \caption{Heatmap of sibling prefixes CIDR size \% for SP-Tuner with /28-/96 thresholds.}
    \label{fig:sp_prefix_size_28_96_p}
  \endminipage
  \hfill
\end{figure*}

In \Cref{fig:sp_domain_category_p} and \Cref{fig:sp_domain_category_24_48_p}, we illustrate the percentage of sibling prefixes based on the number of domains for the default and the /24-/48 thresholds of SP-Tuner algorithm. Finally, we end the appendix of the paper by showing the subnet/CIDR sizes of sibling prefixes for the SP-Tuner algorithm with /24-/48 thresholds in \Cref{fig:sp_prefix_size_24_48_p} and for /28-/96 thresholds in \Cref{fig:sp_prefix_size_28_96_p}.

\end{document}